\documentclass[letterpaper]{article}
\usepackage[T1]{fontenc}
\usepackage{geometry}
\usepackage{setspace}
\usepackage[articletitle=true]{achemso}
\usepackage{graphicx}
\usepackage{float}
\usepackage{amsmath,amssymb}
\usepackage{booktabs}
\usepackage{changepage}
\usepackage{authblk}
\usepackage[hidelinks]{hyperref}
\usepackage{siunitx}
\usepackage{placeins}
\usepackage{needspace}
\usepackage{xurl}
\newlength{\extralength}
\newcommand{\authorcontributions}[1]{\section*{Author Contributions}#1}
\newcommand{\funding}[1]{\section*{Funding}#1}
\newcommand{\dataavailability}[1]{\section*{Data Availability Statement}#1}
\newcommand{\institutionalreview}[1]{\section*{Institutional Review Board Statement}#1}
\newcommand{\informedconsent}[1]{\section*{Informed Consent Statement}#1}
\newcommand{\conflictsofinterest}[1]{\section*{Conflict of Interest}#1}
\newcommand{\reftitle}[1]{}

\title{Zeta potential of magnetic-field-controlled silver nanoparticles}

\author[1]{Y. Tawalbeh}
\author[1,2]{M. F. Pereira}
\affil[1]{Department of Physics, Khalifa University of Science and Technology, 127788 Abu Dhabi, United Arab Emirates}
\affil[2]{Institute of Physics, Czech Academy of Sciences, 18221 Prague, Czech Republic}

\date{Correspondence:mauro.pereira@ku.ac.ae}

\newcommand{\ManuscriptAbstract}{\texorpdfstring{Magnetic-field-dependent nanoparticle size does not by itself determine
zeta potential, which also depends on interfacial charge, electrolyte
screening, and the location of the slipping plane. We extend the
sphere-packing description of silver nanoparticles by distinguishing the
structural surface, the effective charge plane, and the hydrodynamic
slipping plane. The model uses structural radii of \SI{20}{nm} and
\SI{5}{nm} at zero field and \SI{250}{mT}, with reported zeta
potentials of \SI{-13.2}{mV} and \SI{-10.0}{mV}. Effective charge is
related to the defective-shell atom count. For field-independent
interfacial parameters, the linear response without additional charge
weighting decreases with decreasing radius. Its normalized form depends
on three electrokinetic lengths through only two sums, preventing their
independent determination from a field-dependent zeta curve. At
\((\kappa^{-1},t_{\mathrm{lig}},\delta_{\mathrm{slip}})=(3,0,1)\,\mathrm{nm}\),
nonlinear Poisson--Boltzmann electrostatics changes the fitted shell
coefficient from 0.316 to 0.299. Constant total charge still predicts
an increase in zeta magnitude, opposite to the observed trend in the
evaluated scenarios. Independently calibrated shell and area charge laws
differ by less than \SI{0.05}{mV}, so the endpoint agreement does not
distinguish these mechanisms. Propagating the initial size frequencies
gives a narrower endpoint size distribution than measured. With charge
parameters determined from the measured size populations, its predicted
mean zeta potential is \SI{-9.131}{mV}, with standard deviation
\SI{0.994}{mV}. Imposing the measured mean by charge recalibration
does not remove the discrepancy in size width. The combined structural
and electrokinetic model thus identifies both the size-dependent
contribution to zeta potential and the interfacial information needed
to interpret it.}{Structural and electrokinetic model of magnetic-field-controlled silver nanoparticles}}
\newcommand{\ManuscriptKeywords}{silver nanoparticles; zeta potential; magnetic field;
sphere packing; Poisson--Boltzmann equation; slipping plane; identifiability}

\begin{document}
\maketitle

\begin{abstract}
\ManuscriptAbstract
\end{abstract}

\section*{Keywords}
\ManuscriptKeywords

\section{Introduction}
Nanoparticles provide a versatile means of tailoring material properties through control of size, morphology, and surface structure. At these reduced length scales, changes in electronic structure and the increasing proportion of atoms near the surface give rise to physical and chemical responses that differ from those of bulk materials. These features underpin applications in electronic and photonic devices \cite{henini2011handbook, Razeghi2010}, functional nanomaterials \cite{MISHRA2018631}, sensing and biomedical technologies \cite{D4MA90092H, Fabrizio12, Dong:20}, and renewable-energy processes \cite{D1MA00538C}. Reliable control of particle size and its distribution is therefore an important requirement for obtaining reproducible properties across these applications.
The relevance of nanoscale material control also extends to photonics and optoelectronics. Alongside advances in silicon-based platforms for guiding light and controlling polarization \cite{Zafar2025-dl,Zafarpol4,Zafarreview,Zafarpol6,zafar2023band,zafar2023compact,ZafarIEEEAccess,Zafar:22},semiconductor nanoparticles offer size-dependent optical functionality, including quantum-dot photon–emitter interfaces \cite{Uppu2021}. Their optical properties contribute to a broader landscape of semiconductor approaches that includes superlattices and quantum-well structures \cite{ApostolakisPereira+2020+3941+3952, SCHMIELAU2009869,Pereira2020-ii,Oriaku:17,Vaks2022-uk,ma11010002}.
For silver nanoparticles, size and the surrounding dielectric
environment affect the optical response \cite{Kelly2003}, and controlled
size series also show size-dependent antibacterial activity
\cite{Agnihotri2014}. These dependencies motivate methods that control
both the characteristic size and the width of the size distribution
\cite{Tawalbeh2025,Tawalbeh2026APB}. 
{
For particles dispersed in a liquid, however, structural size alone is insufficient to describe their behavior. The charge associated with the interface and the screening produced by mobile ions also influence the interactions between particles \cite{Hunter1981,Ohshima2006}. Relating size selection to electrokinetic properties therefore requires a description of both the particle structure and the surrounding electrolyte.
In addition to electrostatic interactions, the chemical composition of the particle interface can play an important role in colloidal stability and aggregation. 
For example, the formation of a protein corona around silver nanoparticles was reported to improve their colloidal stability and suppress aggregation over 75 days \cite{microorganisms14010178}.}
{
Experimental studies of functional nanomaterials further illustrate the importance  of interfacial chemistry in determining their behavior in liquid environments. Popat and co-workers demonstrated that the functionalization of mesoporous silica nanoparticles with a positively charged chitosan shell produces a pH-responsive system whose behavior depends on the chemical state of the surface coating \cite{Popat2012Chitosan}.
Related investigations of functionalized gold nanoparticles by Patra and co-workers have shown the relevance of electrostatic interactions and surface characterization to biomolecular attachment \cite{Buznyk2020Gold}. In graphene-oxide membranes, Joshi and co-workers examined size-dependent ion adsorption, illustrating the role of the surrounding ionic environment in interfacial processes \cite{Jin2021IonAdsorption}. More recently, work involving Kaushik and Mishra demonstrated the use of zeta-potential characterization alongside particle-size measurements in the development of functionalized, pH-sensitive nanocarriers \cite{Pandey2025Liposomes}. Although these systems differ in composition and application, they collectively illustrate why structural dimensions alone cannot fully characterize the particle--liquid interface. This distinction motivates the separate treatment of structural size, effective charge, and electrokinetic boundary conditions adopted in the present work.}

Magnetic-field-assisted preparation offers a means of varying nanoparticle
size through an external experimental parameter \cite{Kthiri2021,Tawalbeh2026APB}. A theoretical treatment
must connect this parameter to the energetic conditions for particle
formation. Classical nucleation theory provides such a starting point:
the formation of a new phase is governed by the competition between the
free-energy cost of creating an interface and the free-energy decrease
associated with the new phase. Their balance determines the nucleation
barrier and its critical size \cite{Kashchiev2000}. For small particles, the conversion between
size and atom count is also important, since a uniform bulk-density
description does not distinguish the particle interior from its
less efficiently packed surface region \cite{Tawalbeh2025,Tawalbeh2026}.

In our sphere-packing treatment, we described the nanoparticle as a densely
packed core surrounded by a defective shell and used the resulting atom
count in the work of formation \cite{Tawalbeh2025}. Including the magnetic
contribution produced an analytical relation between particle radius and
field strength. For the silver preparations considered there, two
characteristic size points determine the structural trajectory. The same
relation also connects different initial radii to a family of
field-dependent sizes. It therefore provides a basis for examining the
response of a size distribution, in addition to that of a single
characteristic particle.

We subsequently applied this description to silver nanoparticles prepared
under continuous stirring, comparing the parallel and perpendicular
field configurations used in those experiments
\cite{Tawalbeh2026APB}. The analysis related the measured size trends and
their bounds to the structural model and examined the role of experimental
orientation. In a further development, we extended the nucleation
description across magnetic classes by including induced magnetization
and the orientational contribution of collective permanent moments
\cite{Tawalbeh2026}. Magnetite and nickel provided examples of the broader
formulation, while the susceptibility-only limit recovered the analytical
silver-particle relation. These studies connect magnetic contributions,
atomic packing, and size selection, and emphasize the importance of
considering the range of particle sizes. Relating a critical-nucleus scale
to a measured particle population nevertheless requires assumptions about
subsequent growth and aggregation \cite{Tawalbeh2026}.

The structural theory does not determine the electrical potential at the
particle--liquid interface. In particular, zeta potential is defined at the
hydrodynamic slipping plane, which need not coincide with either the
structural surface or the effective charge-bearing surface
\cite{Hunter1981,Delgado2005}. Its experimental value is generally inferred
from electrophoretic mobility through an electrokinetic model
\cite{Delgado2005,Bhattacharjee2016}. Charge distribution, electrolyte
screening, adsorbed molecules, and the location of the slipping plane all
enter this interpretation \cite{Delgado2005,Ohshima2006}. A nanometre-scale displacement between these
boundaries can be significant when the particle itself is only a few
nanometres in radius.
{
The interpretation of measured zeta potentials also depends on the electrokinetic model used to convert electrophoretic mobility into potential, and zeta-potential magnitude alone does not uniquely determine colloidal stability \cite{10.1021/acs.langmuir.1c02056}.}
Consequently, a size-field relation must be combined
with a charge law and an electrostatic solution before it can predict
zeta potential.

{
Experimental studies of silver nanoparticle suspensions have further demonstrated that pH, ionic strength, and electrolyte composition influence surface charge and aggregation, emphasizing that electrokinetic behavior cannot be inferred from particle size alone \cite{10.1021/es902240k}.}

{
The silver nanoparticle preparations reported by Kthiri et al. provide both size information and zeta potentials for this comparison. [34] Particles were synthesized using supernatants from yeast cultures exposed to zero additional field or \SI{250}{mT}; the reported central zeta potentials were \SI{-13.2}{mV} and \SI{-10.0}{mV}, respectively. The field thus labels the preparation history rather than a field applied during electrophoretic measurement. Related investigations examined the combined effects of silver nanoparticles and static magnetic fields on yeast, demonstrating changes in oxidative stress and antioxidant responses \cite{Kthiri2024-us}.These findings further illustrate the complexity of the biological preparation environment, although they do not directly determine the electrokinetic properties of the resulting nanoparticles.}

Here we extend the sphere-packing model to zeta potential by distinguishing
the structural radius \(r\), the effective charge-plane radius \(R_s\),
and the slipping-plane radius \(R_s+\delta_{\mathrm{slip}}\). We relate
effective charge to the defective-shell population and compare this
assumption with charge proportional to surface area. The spherical
Poisson--Boltzmann solution then connects each charge law to the potential
at the slipping plane. We examine the response to decreasing radius,
the interfacial parameter combinations that can be determined from
normalized potentials, and the effect of electrostatic nonlinearity.
Finally, we propagate the initial size frequencies through the structural
relation and calculate the corresponding potential distributions.
Comparison with the measured endpoint sizes separates the consequences
of the structural population model from those of charge calibration.
This combined treatment addresses how much of the electrokinetic response
can be explained by size selection and what further information is needed
to distinguish the interfacial contributions.

\section{Materials and Methods}

\subsection{Structural inputs and sphere-packing model}

\begin{figure}[!htbp]
\begin{adjustwidth}{-\extralength}{0cm}
\centering
\includegraphics[width=\linewidth]{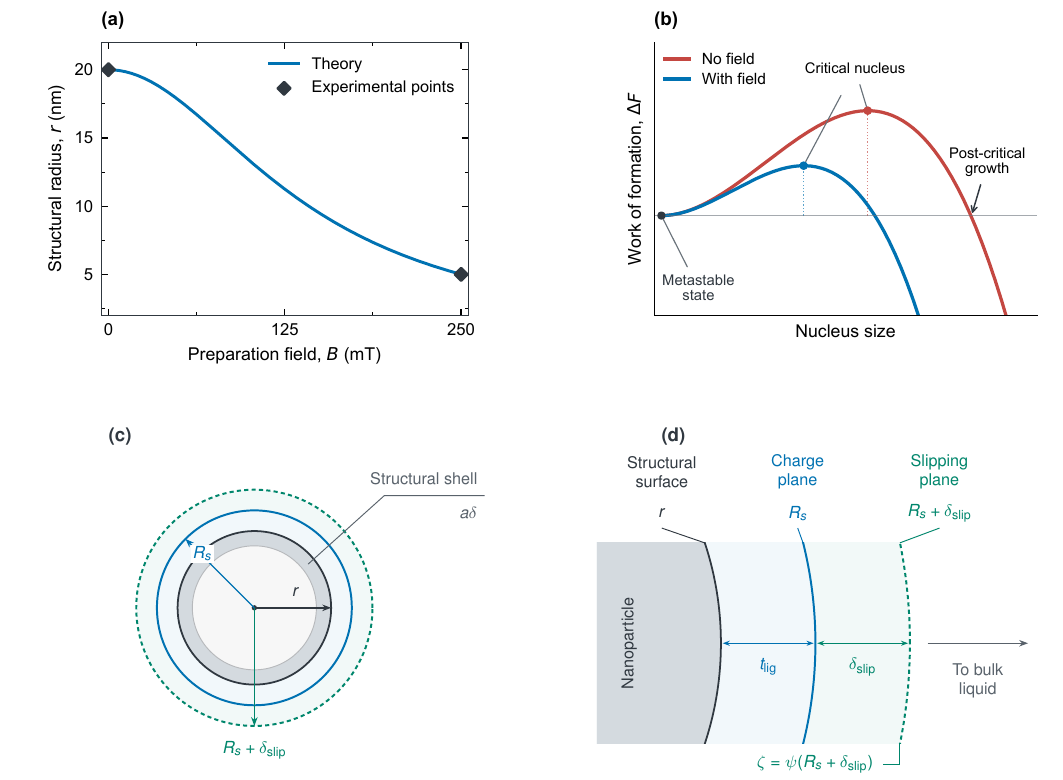}
\end{adjustwidth}
\caption{Nucleation, structural response, and electrokinetic boundaries.
(\textbf{a}) The theoretical radius--field curve passes through the
experimental endpoints, shown as diamonds.
(\textbf{b}) Schematic nucleation work \(\Delta F\), following
Refs.~\cite{Tawalbeh2025,Tawalbeh2026}. The red zero-field and blue
field-prepared curves illustrate the reduction of the barrier and critical
size. Filled circles mark the maxima, dotted vertical lines locate the
critical sizes, and the growth annotation identifies the descending branch.
The grey horizontal line is zero formation work; the leftmost dark point
denotes the metastable parent state.
(\textbf{c,d}) The black structural surface has radius \(r\), the blue
charge plane has radius \(R_s=r+t_{\mathrm{lig}}\), and the dashed
green slipping plane has radius \(R_s+\delta_{\mathrm{slip}}\).
The grey annulus represents the defective structural shell of thickness
\(a\delta\). Blue and green tinted regions distinguish the effective
offsets, not separately identified material layers. Radius arrows in
(\textbf{c}) originate at the particle center; thickness arrows in
(\textbf{d}) span the corresponding radial intervals. The potential
\(\zeta=\psi(R_s+\delta_{\mathrm{slip}})\) is evaluated at the green
boundary. Panels (\textbf{b--d}) are schematic: the free-energy curves
are not quantitative fits, and interfacial offsets are not to scale.}
\label{fig:model}
\end{figure}

We begin with the classical nucleation picture used in our preceding
sphere-packing work \cite{Tawalbeh2025,Tawalbeh2026APB,Tawalbeh2026}. Nucleation describes
the formation of a new phase from a metastable parent phase. Metastability
refers to a state in which the system resides temporarily at a local
free-energy minimum, rather than the global minimum associated with the
stable phase. To leave this state, the system must overcome a free-energy
barrier. We describe the reversible work required to form a nucleus of a
given size by the work of formation, \(\Delta F\). Its maximum along the
nucleation coordinate is the barrier that separates subcritical clusters
from post-critical growth \cite{Kashchiev2000}, as illustrated in Figure~\ref{fig:model}b.

The barrier arises from two competing contributions. Creating an interface
between the nucleus and its surroundings costs free energy, whereas
forming the new phase lowers the bulk free energy when the chemical
driving force favors that phase. For a spherical nucleus, the interfacial
cost grows with its area, while the bulk contribution grows with its atom
count. The surface cost is therefore particularly important for small
clusters. Below the critical size, a cluster can lower its work of
formation by shrinking; beyond the critical size, further growth lowers
that work. The critical nucleus is consequently located at the barrier
maximum, not at a stable equilibrium minimum. This distinction allows us
to use the stationarity condition to determine a critical size without
interpreting it as a prediction of an arrested final particle size.

In the present application, the magnetic field \(B\) labels the
preparation condition associated with a particle population; it is not a
field applied during electrophoretic measurement. The reported central
values are \(\zeta_0=\SI{-13.2}{mV}\) at \(B=0\) and
\(\zeta_B=\SI{-10.0}{mV}\) at \(B_2=\SI{250}{mT}\)
\cite{Kthiri2021}, giving
\(\rho_{\mathrm{obs}}=|\zeta_B|/|\zeta_0|=0.758\). Since no uncertainty
was reported for this pair, we treat it as a deterministic endpoint
constraint rather than a statistical estimate.

To retain the structural description of the preceding sphere-packing
theory \cite{Tawalbeh2025}, we take
\[
r_1=r(0)=\SI{20}{nm},\qquad r_2=r(B_2)=\SI{5}{nm}
\]
We keep these inputs fixed in the characteristic-particle benchmark. The
population calculation introduced below is a separate extension and does
not replace them with empirical means. We use the dimensionless size
\(x=r/a\), with effective atomic radius
\(a=\SI{0.144}{nm}\) and dimensionless structural shell thickness
\(\delta=4\). The model atom count and its shell contribution are
\begin{align}
n(x)&=\phi_{\mathrm b}(x-\delta)^3+
\phi_{\mathrm d}\left[x^3-(x-\delta)^3\right]\nonumber\\
n_{\mathrm{shell}}(x)&=\phi_{\mathrm d}\left[x^3-(x-\delta)^3\right]
\label{eq:packing}
\end{align}
Here \(\phi_{\mathrm b}=\pi/(3\sqrt{2})\) represents the maximum density
for congruent sphere packing \cite{Hales2005}, and
\(\phi_{\mathrm d}=0.5172879\) is the defective-shell
parameter \cite{Tawalbeh2025}. This construction represents a densely
packed core surrounded by a shell of reduced packing efficiency. It
connects the atomic assembly to the bulk contribution in the work of
formation, rather than replacing the atom count by a uniform-density
volume at every size. Thus \(n(x)\) and \(n_{\mathrm{shell}}(x)\)
retain their meaning as model atom counts in the packing
model. They are effective counts rather than direct atomic measurements. All evaluated
radii satisfy \(x>\delta\). The physical thickness
\(a\delta=\SI{0.576}{nm}\) belongs to the structural shell and is
distinct from the electrokinetic offsets introduced below.

The spherical work of formation combines the classical surface
and bulk contributions with the field-associated driving term
\cite{Tawalbeh2025}. We take the orientation factor as
unity, corresponding to the deterministic-orientation case:
\begin{equation}
\Delta F(x,B)=-n(x)\left[\Delta\mu+E_B(B/B_2)^2\right]+E_sx^2
\label{eq:formation}
\end{equation}
Here \(\Delta\mu\) is the positive chemical driving energy per atom for
formation of the new phase. The coefficient \(E_s=4\pi\gamma a^2\)
expresses the surface term through the interfacial free energy per area
\(\gamma\), so that \(E_sx^2=4\pi\gamma r^2\). The coefficient
\(E_B\) is the field-associated driving energy per atom at \(B_2\);
its definition is given in Supporting Information Section S2.
For positive \(E_B\), the field-dependent term lowers the work of
formation at a given size. In this structural model it lowers the barrier
and shifts its maximum toward a smaller nucleus, as the schematic in
Figure~\ref{fig:model}b illustrates. The sketch is not a fitted
free-energy curve and does not add a finite-radius equilibrium minimum
to Equation~\eqref{eq:formation}.

We obtain the critical size by applying
\(\partial\Delta F/\partial x=0\), which balances the size derivatives
of the interfacial cost and the driving contribution:
\begin{equation}
2E_s\frac{x(B)}{n'[x(B)]}=\Delta\mu+E_B(B/B_2)^2
\label{eq:stationarity}
\end{equation}
The relevant stationary branch has
\(\partial^2\Delta F/\partial x^2<0\), as required for the barrier
maximum. Eliminating the energy coefficients with the two structural
inputs gives
\begin{equation}
\frac{x(B)}{n'[x(B)]}=\frac{x_1}{n'(x_1)}+
\left[\frac{x_2}{n'(x_2)}-\frac{x_1}{n'(x_1)}\right](B/B_2)^2,
\qquad x_1=r_1/a,\quad x_2=r_2/a
\label{eq:trajectory}
\end{equation}
For \(0\leq B\leq B_2\), we select the larger solution with
\(x>\delta\); Supporting Information Section S2 gives the explicit
branch equation and endpoint checks. Figure~\ref{fig:model}a shows the
resulting trajectory. It follows from the work of formation and the two
endpoint constraints, rather than an additional empirical interpolation.

The critical nucleus provides an initial post-critical size scale. As in
the preceding framework \cite{Tawalbeh2026}, relating this scale to
measured particle sizes assumes that subsequent growth, coalescence,
aggregation, and ripening do not reverse the size ordering established
during nucleation. We use the structural trajectory under this assumption, with later
growth processes outside the present calculation. Moreover, applying the
trajectory to preparation-labelled data does not establish direct magnetic
action during particle formation. The structural model determines the
size input to the electrokinetic calculation, but it does not determine
the particle charge.

\Needspace{7\baselineskip}
\subsection{Three-radius electrokinetic model}

The structural trajectory supplies a particle radius, but zeta potential
is defined at a hydrodynamic boundary in the surrounding liquid. We must
therefore specify both where effective charge is represented and where
the potential is evaluated. As shown in Figure~\ref{fig:model}c,d, we
distinguish these boundaries through
\begin{equation}
R_s(B)=r(B)+t_{\mathrm{lig}}(B),\qquad
\zeta(B)=\psi\!\left[R_s(B)+\delta_{\mathrm{slip}}(B)\right]
\label{eq:radii}
\end{equation}
The radius \(R_s\) retains its charge-bearing meaning, while
\(t_{\mathrm{lig}}\) locates this effective plane relative to the
structural surface. It can represent ligand, corona, and solvation
contributions and is not identified with a measured ligand thickness.
The offset \(\delta_{\mathrm{slip}}\) locates the slipping plane
outside the charge plane. Liquid inside the slipping boundary is treated
as moving with the particle in the electrokinetic interpretation; zeta
potential is the potential at that boundary relative to the bulk liquid
\cite{Hunter1981,Delgado2005}. Neither offset is independently inferred
from the two endpoint measurements.

To obtain this potential, we solve the electrostatic problem outside a
sharp spherical charge plane. The surrounding mobile ions screen the
particle charge, and their response is represented by the
Poisson--Boltzmann description. We assume spherical symmetry, uniform
permittivity \(\varepsilon\), and a symmetric monovalent electrolyte.
Here \(e\) is the elementary charge, \(k_{\mathrm B}\) is Boltzmann's
constant, and \(T\) is the temperature. When
\(|e\psi|/(k_{\mathrm B}T)\ll1\), linearizing the ionic response
gives the linear spherical Poisson--Boltzmann equation
\cite{Hunter1981,Ohshima2006}. Here \(R\) denotes radial position in
the liquid, measured from the particle center, rather than the fixed
structural radius \(r\). At each preparation condition, the equation and
boundary conditions are
\begin{align}
\frac{1}{R^2}\frac{d}{dR}\left(R^2\frac{d\psi}{dR}\right)
&=\kappa^2\psi,\qquad R\geq R_s\nonumber\\
-\varepsilon\left.\frac{d\psi}{dR}\right|_{R_s^+}
&=\frac{Q_{\mathrm{eff}}}{4\pi R_s^2}
=\sigma_{\mathrm{eff}},\qquad \psi(\infty)=0
\label{eq:linear_pb_boundary}
\end{align}
The Debye parameter \(\kappa\) specifies the screening, and
\(\kappa^{-1}\) is the screening length. The notation \(R_s^+\)
denotes the liquid side of the charge plane. The liquid-side boundary
condition relates the radial electric field to the total effective
charge. Thus \(Q_{\mathrm{eff}}\) and
\(\sigma_{\mathrm{eff}}=Q_{\mathrm{eff}}/(4\pi R_s^2)\) describe
the same boundary charge, expressed as a total or per charge-plane area;
they are not independent fit parameters. This effective charge is not a
measurement of bare metallic charge or of the particle together with all
surrounding counterions. The condition at infinity fixes the bulk
potential as the reference zero.

\begin{figure}[!htbp]
\begin{adjustwidth}{-\extralength}{0cm}
\centering
\includegraphics[width=\linewidth]{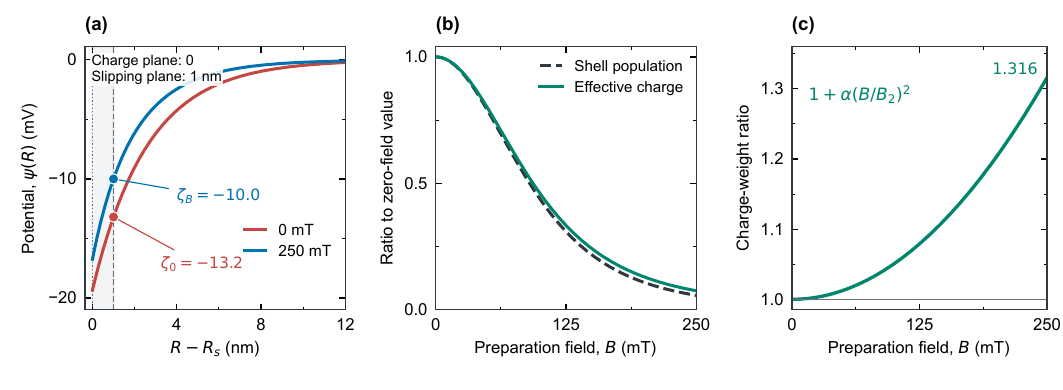}
\end{adjustwidth}
\caption{Electrostatic potential and shell-dependent charge.
(\textbf{a}) Linear Poisson--Boltzmann profiles for the two characteristic
particles: red at zero field and blue at \SI{250}{mT}. The dotted vertical
line locates the charge plane at \(R-R_s=0\); the dashed line locates
the slipping plane at \(R-R_s=\delta_{\mathrm{slip}}=\SI{1}{nm}\).
The pale region lies between these planes. Circles at the slipping plane
mark the calibrated potentials \SI{-13.2}{mV} and \SI{-10.0}{mV}.
(\textbf{b}) The grey dashed curve is
\(n_{\mathrm{shell}}[x(B)]/n_{\mathrm{shell}}(x_1)\); the green solid
curve is \(|Q_{\mathrm{eff}}(B)|/|Q_{\mathrm{eff}}(0)|\).
(\textbf{c}) The charge weight per shell count relative to its zero-field
value is \(1+\alpha f(B)\). The endpoint annotation 1.316 is
\(1+\alpha\), not the total-charge ratio.
All panels use
\((\kappa^{-1},t_{\mathrm{lig}},\delta_{\mathrm{slip}})=(3,0,1)\,\mathrm{nm}\),
\(f(B)=(B/B_2)^2\), and \(\alpha=0.316017\). Thus the charge weight
increases while the total effective-charge magnitude decreases.}
\label{fig:electrokinetic}
\end{figure}

The decaying spherical solution has a radial dependence proportional to
\(e^{-\kappa R}/R\). Expressing its amplitude through the charge-plane
potential gives
\begin{equation}
\psi(R)=\psi(R_s)\frac{R_s}{R}\exp[-\kappa(R-R_s)]
\label{eq:radial_decay}
\end{equation}
Differentiation at the charge plane gives
\(\left.d\psi/dR\right|_{R_s^+}=-(\kappa+1/R_s)\psi(R_s)\).
Substitution in the charge boundary condition then determines the
amplitude and hence the full potential:
\begin{align}
\psi(R_s)&=\frac{Q_{\mathrm{eff}}}
{4\pi\varepsilon R_s(1+\kappa R_s)}\nonumber\\
\psi(R)&=\frac{Q_{\mathrm{eff}}}
{4\pi\varepsilon R(1+\kappa R_s)}\exp[-\kappa(R-R_s)]
\label{eq:radial_potential}
\end{align}
This expression separates the spherical geometric decay from the
exponential screening. The factor \(1+\kappa R_s\) follows from
applying the charge boundary condition to that screened spherical
solution. Evaluating it at the slipping plane,
\(R=R_s+\delta_{\mathrm{slip}}\), gives the surface-density and
total-charge forms of zeta potential:
\begin{align}
\zeta(B)&=\frac{\sigma_{\mathrm{eff}}(B)R_s^2(B)}
{\varepsilon(B)[R_s(B)+\delta_{\mathrm{slip}}(B)][1+\kappa(B)R_s(B)]}
e^{-\kappa(B)\delta_{\mathrm{slip}}(B)}\nonumber\\
&=\frac{Q_{\mathrm{eff}}(B)}
{4\pi\varepsilon(B)[R_s(B)+\delta_{\mathrm{slip}}(B)][1+\kappa(B)R_s(B)]}
e^{-\kappa(B)\delta_{\mathrm{slip}}(B)}
\label{eq:zeta}
\end{align}
Zeta potential therefore need not equal the potential at the charge
plane, even for a perfectly spherical particle. For a negative charge,
the potential remains negative but approaches zero as the distance into
the liquid increases, as illustrated in Figure~\ref{fig:electrokinetic}a.
When \(\delta_{\mathrm{slip}}=0\), the two evaluation planes coincide;
otherwise the slipping-plane value is attenuated by both geometry and
screening. Equation~\eqref{eq:zeta} is a charge-to-potential map, not a
complete electrophoretic-mobility theory. Supporting Information
Section S3 gives the corresponding derivation and attenuation relation.

The electrostatic calculation determines the potential for a specified
charge; it does not by itself determine how the surface acquires that
charge. We therefore introduce a separate constitutive hypothesis. We
take the defective-shell population as the primary structural quantity
because it represents the near-surface part of the packing model. At
fixed \(\delta\), its leading large-particle dependence is proportional
to surface area, while the same geometry supplies finite-size
corrections. This provides a physically motivated reference without
assuming that every shell atom is accessible to the solvent or carries
the same microscopic charge. In particular, \(a\delta\) is a
structural thickness, not an electrostatic penetration length. We write
\begin{equation}
Q_{\mathrm{eff}}(B)=q_0[1+\alpha f(B)]
n_{\mathrm{shell}}[x(B)],\qquad f(0)=0,\quad f(B_2)=1
\label{eq:charge}
\end{equation}
The zero-field coefficient \(q_0\) is an effective charge weight per
model shell count and carries the charge sign. It is not an assertion
that an individual silver atom has this microscopic charge. Since
\(f(0)=0\), the zero-field relation is
\(q_0=Q_{\mathrm{eff}}(0)/n_{\mathrm{shell}}(x_1)\). Inverting
Equation~\eqref{eq:zeta} at zero field makes its calibration explicit:
\begin{equation}
q_0=\frac{4\pi\varepsilon(0)[R_s(0)+\delta_{\mathrm{slip}}(0)]
[1+\kappa(0)R_s(0)]e^{\kappa(0)\delta_{\mathrm{slip}}(0)}\zeta_0}
{n_{\mathrm{shell}}(x_1)}
\label{eq:q0_calibration}
\end{equation}
Thus the assumed interface and the zero-field potential determine the zero-field
charge weight. They do not identify its chemical origin. At the second
endpoint, \(f(B_2)=1\), so the effective weight becomes
\(q_0(1+\alpha)\). The dimensionless coefficient \(\alpha\) is
therefore the fractional endpoint change of this weight:
\begin{equation}
\alpha=\frac{Q_{\mathrm{eff}}(B_2)/n_{\mathrm{shell}}(x_2)}
{Q_{\mathrm{eff}}(0)/n_{\mathrm{shell}}(x_1)}-1
\label{eq:alpha_meaning}
\end{equation}
More generally, the assumed charge law separates the change in weight
from the change in the number of shell contributions:
\begin{align}
\frac{Q_{\mathrm{eff}}(B)/n_{\mathrm{shell}}[x(B)]}
{Q_{\mathrm{eff}}(0)/n_{\mathrm{shell}}(x_1)}
&=1+\alpha f(B)\nonumber\\
\frac{Q_{\mathrm{eff}}(B)}{Q_{\mathrm{eff}}(0)}
&=[1+\alpha f(B)]\frac{n_{\mathrm{shell}}[x(B)]}{n_{\mathrm{shell}}(x_1)}
\label{eq:weight_and_charge}
\end{align}
This distinction is important when interpreting a positive \(\alpha\).
For a fixed charge sign, it means a larger magnitude of effective charge
weight at the field endpoint, not necessarily a larger total
effective-charge magnitude. For example, at the illustrative interface
\((\kappa^{-1},t_{\mathrm{lig}},\delta_{\mathrm{slip}})=(3,0,1)\,\mathrm{nm}\),
the linear endpoint calibration gives \(\alpha=0.316017\). The weight
magnitude increases by approximately 31.6\%, but the shell-count ratio
is only 0.0572083, giving a total effective-charge magnitude ratio of
0.075287. Figure~\ref{fig:electrokinetic}b,c separates these two changes.
An unamplified response, \(\alpha=0\), means an unchanged charge weight,
not an uncharged particle or an unchanged total charge.

Changing \(\alpha\) does not change \(\phi_{\mathrm d}\),
\(\delta\), or the structural trajectory. A packing-fraction ceiling
therefore does not bound this coefficient. We require
\(1+\alpha f(B)>0\) to retain the charge sign; for the response shapes
used here, with \(0\leq f(B)\leq1\), this requires \(\alpha>-1\).
The function \(f(B)\) specifies how the weight changes between the two
conditions, separately from its endpoint amplitude. We use
\(f(B)=(B/B_2)^2\) as the leading even, analytic response,
and compare it with linear-in-\(|B|\) and early-saturating alternatives.
Both the shape and the charge law are model assumptions. In particular,
\(\alpha\) summarizes the charge-weight change associated with the
preparations under a specified interface; it is not a directly measured
magnetic charging constant.

To examine what follows from the choice of charge law, we also use the
following unamplified closures:
\begin{equation}
\frac{Q_{\mathrm{eff}}(B)}{Q_{\mathrm{eff}}(0)}=\begin{cases}
1,&\text{constant total effective charge}\\
[R_s(B)/R_s(0)]^2,&\text{constant charge-plane density}\\
n_{\mathrm{shell}}[x(B)]/n_{\mathrm{shell}}(x_1),&\text{unamplified shell}
\end{cases}
\label{eq:benchmark_closures}
\end{equation}
For comparison, we allow the area and shell charge laws to reproduce
the same endpoint potentials by
assigning each its own coefficient multiplying the same quadratic
response. Their zero-field amplitudes and endpoint coefficients are
calibrated separately within each electrostatic model. The two endpoints
are consequently matched by construction; only the interior predictions
can distinguish the two charge laws at a fixed interface.

\subsection{Ratio analysis and applicability}

To separate the relative response from its zero-field amplitude, we define
\(\rho(B)=|\zeta(B)|/|\zeta_0|\), the zeta magnitude normalized by its
zero-field value. Applying Equation~\eqref{eq:zeta} at the two conditions
gives
\begin{align}
\rho(B)&=\left|\frac{Q_{\mathrm{eff}}(B)}{Q_{\mathrm{eff}}(0)}\right|
\frac{\varepsilon(0)}{\varepsilon(B)}
\frac{R_s(0)+\delta_{\mathrm{slip}}(0)}{R_s(B)+\delta_{\mathrm{slip}}(B)}
\frac{1+\kappa(0)R_s(0)}{1+\kappa(B)R_s(B)}\nonumber\\
&\quad\times
\exp[\kappa(0)\delta_{\mathrm{slip}}(0)-\kappa(B)\delta_{\mathrm{slip}}(B)]
\label{eq:ratio}
\end{align}
Both reported potentials are negative, so this ratio describes their
relative magnitude without modeling charge reversal. In this
calculation we hold
\(\varepsilon,\kappa^{-1},t_{\mathrm{lig}},\delta_{\mathrm{slip}}\)
constant between preparations. Substituting the shell hypothesis then
gives the normalized response and the endpoint value of \(\alpha\):
\begin{align}
\rho(B)&=[1+\alpha f(B)]
\frac{n_{\mathrm{shell}}[x(B)]}{n_{\mathrm{shell}}(x_1)}
\frac{r_1+t_{\mathrm{lig}}+\delta_{\mathrm{slip}}}{r(B)+t_{\mathrm{lig}}+\delta_{\mathrm{slip}}}
\frac{1+\kappa(r_1+t_{\mathrm{lig}})}{1+\kappa[r(B)+t_{\mathrm{lig}}]}
\label{eq:transfer}\\
\alpha&=\rho_{\mathrm{obs}}
\frac{n_{\mathrm{shell}}(x_1)}{n_{\mathrm{shell}}(x_2)}
\frac{r_2+t_{\mathrm{lig}}+\delta_{\mathrm{slip}}}{r_1+t_{\mathrm{lig}}+\delta_{\mathrm{slip}}}
\frac{1+\kappa(r_2+t_{\mathrm{lig}})}{1+\kappa(r_1+t_{\mathrm{lig}})}-1
\label{eq:alpha_inverse}
\end{align}
The common charge amplitude and defective-shell factor cancel from the
ratio. Equation~\eqref{eq:alpha_inverse} also shows why an endpoint fit
cannot validate the shell law: a positive observed ratio always supplies a
positive endpoint multiplier, conditional on the chosen interface.

The applicability of the linear map must be checked at the charge plane,
where the potential can exceed the value at the slipping plane. Using the
linear solution to recover this potential gives
\begin{equation}
\left|\frac{e\psi(R_s)}{k_{\mathrm B}T}\right|
=\frac{e|\zeta|}{k_{\mathrm B}T}\frac{R_s+\delta_{\mathrm{slip}}}{R_s}
\exp(\kappa\delta_{\mathrm{slip}})
\label{eq:linearity}
\end{equation}
This dimensionless quantity compares electrostatic and thermal energies,
with elementary charge \(e\), Boltzmann's constant \(k_{\mathrm B}\),
and temperature \(T\). For monovalent ions the weak-potential condition is
\(|e\psi(R_s)|/(k_{\mathrm B}T)\ll1\). Because the back-propagation
itself uses the linear model, this diagnostic is not an error estimate.
The inverse maps mark regions where either endpoint has
\(|e\psi(R_s)|/(k_{\mathrm B}T)\geq1\). Values below this threshold
do not automatically establish quantitative accuracy, which motivates the
nonlinear comparison described below.

\Needspace{8\baselineskip}
\subsection{Size-dependent population response}

The population calculation extends the structural relation in
Section 2.1 to the digitized zero-field size distribution
\cite{Kthiri2021,Tawalbeh2025}. For preparation \(j=1,2\), the
frequency \(f_{ji}\) at bin center \(r_{ji}\) defines
\(w_{ji}=f_{ji}/\sum_i f_{ji}\). These normalized frequencies are
interpreted as particle-number fractions. The population mean of an
observable is therefore its frequency-weighted average, rather than an
equal-weight average over size bins. The structural relation determines
the field-dependent sizes, while the frequencies specify their relative
abundance. The calculation assumes that these initial fractions are
preserved under the structural mapping; it does not describe changes in
particle abundance caused by size-dependent formation or aggregation.

For each initial \(x_{1i}=r_{1i}/a\), the field-dependent size satisfies
\begin{equation}
\frac{x_i(B)}{n'[x_i(B)]}=
\left[1+(\lambda-1)(B/B_2)^2\right]\frac{x_{1i}}{n'(x_{1i})},
\qquad \lambda=\frac{x_2/n'(x_2)}{x_1/n'(x_1)}
\label{eq:population_flow}
\end{equation}
where \(r_i(B)=a x_i(B)\). The characteristic endpoints defined in
Section 2.1 give \(\lambda=4.206878\). The larger real root is
selected, recovering \(r_i(0)=r_{1i}\); its explicit endpoint form
is given in Supporting Information Section S9.5. Each solution retains
its initial weight \(w_{1i}\).

Applying the shell-linked charge law with common interfacial parameters
gives the potential of each size class:
\begin{equation}
\zeta(r,B)=\frac{q_0[1+\alpha_{\mathrm{pop}}f(B)]n_{\mathrm{shell}}(r/a)
e^{-\kappa\delta_{\mathrm{slip}}}}
{4\pi\varepsilon(r+t_{\mathrm{lig}}+\delta_{\mathrm{slip}})
[1+\kappa(r+t_{\mathrm{lig}})]}
\label{eq:population_kernel}
\end{equation}
Here \(f(B)=(B/B_2)^2\). A single zero-field effective charge weight
\(q_0\) and a single field coefficient \(\alpha_{\mathrm{pop}}\)
apply to all size classes. Neither parameter is itself averaged over
particles; the population mean is obtained by averaging the resulting
potentials. The subscript distinguishes this calibration from the
characteristic-particle calculation.

We assume that the reported zeta potentials equal the
number-weighted population means. This identification is an assumption about the measurement,
because the structural relation does not specify the electrophoretic
weighting \cite{Delgado2005}.
The conditions \(\sum_iw_{1i}\zeta(r_{1i},0)=\zeta_0\) and
\(\sum_iw_{2i}\zeta(r_{2i},B_2)=\zeta_B\) determine the two
charge parameters. At zero field, \(f(0)=0\), so the first condition
determines \(q_0\). At \(B_2\), \(f(B_2)=1\), and the charge
weight is multiplied by \(1+\alpha_{\mathrm{pop}}\).
Dividing the two mean potentials eliminates \(q_0\) and the common
electrostatic prefactor. Solving the resulting ratio for
\(\alpha_{\mathrm{pop}}\) gives
\begin{align}
q_0&=\frac{4\pi\varepsilon\zeta_0e^{\kappa\delta_{\mathrm{slip}}}}
{\displaystyle\sum_iw_{1i}
\frac{n_{\mathrm{shell}}(r_{1i}/a)}
{(r_{1i}+t_{\mathrm{lig}}+\delta_{\mathrm{slip}})[1+\kappa(r_{1i}+t_{\mathrm{lig}})]}}
\nonumber\\
\alpha_{\mathrm{pop}}&=\rho_{\mathrm{obs}}
\frac{\displaystyle\sum_iw_{1i}
\frac{n_{\mathrm{shell}}(r_{1i}/a)}
{(r_{1i}+t_{\mathrm{lig}}+\delta_{\mathrm{slip}})[1+\kappa(r_{1i}+t_{\mathrm{lig}})]}}
{\displaystyle\sum_iw_{2i}
\frac{n_{\mathrm{shell}}(r_{2i}/a)}
{(r_{2i}+t_{\mathrm{lig}}+\delta_{\mathrm{slip}})[1+\kappa(r_{2i}+t_{\mathrm{lig}})]}}-1
\label{eq:population_calibration}
\end{align}
The first relation enforces the mean \(\zeta_0=\SI{-13.2}{mV}\)
over the measured zero-field sizes. Individual initial potentials are
then calculated from Equation~\eqref{eq:population_kernel}; they are not
separately fitted measurements. The second relation enforces
\(\zeta_B=\SI{-10.0}{mV}\) over the measured field-prepared sizes,
with their own weights \(w_{2i}\). At the reference interface this gives
\(\alpha_{\mathrm{pop}}=0.196860\).

The structural prediction uses a different endpoint population:
the radii \(r_i(B_2)\) generated from the initial sizes, retaining
the initial weights \(w_{1i}\). Figure~\ref{fig:population} evaluates
the same charge law on that predicted population without refitting.
The two populations coincide at zero field, where the mean is imposed,
but not at \(B_2\). Consequently, the predicted endpoint mean need not
equal the mean imposed on the measured endpoint sizes. This is a
consistency test of structural propagation with the measured-size charge
calibration, not an independent prediction from zero-field data alone.
An endpoint-constrained population model instead
replaces \(r_{2i},w_{2i}\) in the denominator of
Equation~\eqref{eq:population_calibration} by
\(r_i(B_2),w_{1i}\). This gives
\(\alpha_{\mathrm{pop}}=0.310829\), with the reported endpoint
mean satisfied by construction. Figure S2 compares the two calibrations.

Population moments and percentiles are calculated at the original bin
centers. For histogram display, each initial bin is represented by
uniform mass on \([r_{1i}-0.5,r_{1i}+0.5]\,\mathrm{nm}\).
This mass is propagated through the structural and electrokinetic
relations and integrated into shared \SI{0.5}{mV} intervals.
The displayed percentages sum to 100\% at each field. The within-bin
representation is a display assumption and is not used to fit the
charge parameters. The resulting widths and percentile intervals
describe particle-to-particle variation under the common-interface model,
not uncertainty in the population mean.

\subsection{Nonlinear benchmark and numerical verification}

We test the weak-potential approximation by solving the nonlinear
electrostatic problem while retaining the assumed charging law and
boundaries. For the same sharp planes, a symmetric monovalent electrolyte,
and common permittivity, we write the nonlinear equation in the physical
potential \(\psi\). As in the linear derivation, \(R\) measures
distance from the particle center in the liquid, distinct from the
fixed structural radius \(r\):
\begin{align}
\frac{d^2\psi}{dR^2}+\frac{2}{R}\frac{d\psi}{dR}
&=\kappa^2\frac{k_{\mathrm B}T}{e}
\sinh\!\left(\frac{e\psi}{k_{\mathrm B}T}\right),\qquad R\geq R_s
\nonumber\\
-\varepsilon\left.\frac{d\psi}{dR}\right|_{R_s^+}
&=\frac{Q_{\mathrm{eff}}}{4\pi R_s^2},\qquad
\psi(\infty)=0,\qquad \zeta=\psi(R_s+\delta_{\mathrm{slip}})
\label{eq:nonlinear}
\end{align}
We solve the inverse problem at the two observed potentials to determine
the corresponding effective charges within this nonlinear model. The
shell charge law then gives the response coefficient as
\begin{equation}
\alpha_{\mathrm{NL}}=
\frac{Q_{\mathrm{eff}}(B_2)}{Q_{\mathrm{eff}}(0)}
\frac{n_{\mathrm{shell}}(x_1)}{n_{\mathrm{shell}}(x_2)}-1
\label{eq:nonlinear_alpha}
\end{equation}
For the numerical comparison, we use \(T=\SI{298.15}{K}\),
\(\kappa^{-1}=1.5,3,5,8\,\mathrm{nm}\), \(t_{\mathrm{lig}}=0\), and
\(\delta_{\mathrm{slip}}=\SI{1}{nm}\). These screening lengths specify the comparison scenarios;
they have not been measured for the preparations. To examine the
dependence on the assumed interface, the charge-plane inverse maps cover
\(0.5\leq \kappa^{-1}\leq\SI{8}{nm}\) and
\(0\leq t_{\mathrm{lig}}\leq\SI{2.5}{nm}\), at fixed
\(\delta_{\mathrm{slip}}=0,1,2\,\mathrm{nm}\). The separate endpoint
applicability maps use the same screening range and
\(0\leq\delta_{\mathrm{slip}}\leq\SI{4}{nm}\), with
\(t_{\mathrm{lig}}=0\). We sample the reference field curves every
\SI{5}{mT} and additionally evaluate the analytic linear curves on a
finer grid.

We solve the nonlinear boundary-value problem by adaptive collocation
with a remote decaying spherical tail and check inverse recovery from a
prescribed slipping-plane potential using a separate charge-boundary
solve. To assess numerical convergence, we change the outer boundary from
24 to 16 and 32 Debye lengths, double the initial mesh, and tighten the
tolerances tenfold. These changes leave the reference charge ratios and
response coefficients unchanged at the reported precision. In the
\SI{0.01}{mV} weak-potential limit, the recovered charge agrees with the
linear charge to within \(2.9\times10^{-8}\) relative difference.
Supporting Information Section S7 provides the equations, settings, and
convergence results. These checks establish numerical convergence for the specified
electrostatic boundary problem.

\section{Results}

\subsection{Finite-shell correction and direction of the unamplified response}

We first examine how the structural trajectory changes the shell contribution
before introducing any additional charge weighting. The trajectory gives
\(r=\SI{11.2}{nm}\) at \SI{125}{mT}. Between the fixed endpoints,
\(n_{\mathrm{shell}}(x_2)/n_{\mathrm{shell}}(x_1)=0.0572083\), so the shell count
per characteristic particle decreases by a factor of 17.5. Its geometric
fraction nevertheless increases from 0.0839 to 0.307, while its packing-weighted
fraction increases from 0.0602 to 0.237. Thus a larger fraction of the smaller
particle belongs to the shell, although the shell itself contains fewer atoms.

To compare the shell hypothesis with ordinary surface-area scaling, we
rearrange the shell-count expression without introducing an approximation.
For \(r>a\delta\),
\begin{equation}
n_{\mathrm{shell}}(r/a)=\frac{3\phi_{\mathrm d}\delta}{a^2}r^2
\left[1-\frac{a\delta}{r}+\frac{(a\delta)^2}{3r^2}\right]
\label{eq:finite_shell}
\end{equation}
The square-bracketed term gives the finite-shell correction to the
\(r^2\) dependence. Its ratio at \(r_2\) and \(r_1\) is 0.915332, so the
shell ratio differs from the structural-area ratio by 8.47\%. When
\(t_{\mathrm{lig}}=0\), the charge-plane area is the structural area, and
the area closure in Equation~\eqref{eq:benchmark_closures} provides this
comparison directly.

We can also determine the direction of the unamplified response without
selecting a particular interface. Within the field-invariant linear model,
differentiating the physical potential gives
\begin{equation}
\left.\frac{d\ln|\zeta|}{dr}\right|_{\alpha=0}
=\frac{d\ln n_{\mathrm{shell}}(r/a)}{dr}
-\frac{1}{r+t_{\mathrm{lig}}+\delta_{\mathrm{slip}}}
-\frac{1}{r+t_{\mathrm{lig}}+\kappa^{-1}}>0
\quad(r>a\delta)
\label{eq:monotonic}
\end{equation}
For nonnegative offsets and a positive Debye length, the unamplified zeta
magnitude therefore decreases as the radius decreases. At the present
endpoints, the normalized response satisfies
\begin{equation}
\frac{n_{\mathrm{shell}}(x_2)}{n_{\mathrm{shell}}(x_1)}
<\left.\rho(B_2)\right|_{\alpha=0}
<\left(\frac{r_1}{r_2}\right)^2
\frac{n_{\mathrm{shell}}(x_2)}{n_{\mathrm{shell}}(x_1)}=0.915332<1
\label{eq:unamplified_bound}
\end{equation}
Since the reported ratio lies within these bounds, the data do not require
additional charge weighting for every possible interface. For example,
with both offsets set to zero, the unamplified linear shell closure gives
the reported ratio at \(\kappa^{-1}=\SI{1.49}{nm}\), whereas the constant
charge-plane density closure does so at \(\kappa^{-1}=\SI{2.39}{nm}\).
We give the proof and the analytic zero-response boundary in Supporting
Information Section S5.

\subsection{Nonlinear correction and comparison of charge laws}

\begin{figure}[htbp]
\begin{adjustwidth}{-\extralength}{0cm}
\centering
\includegraphics[width=\linewidth]{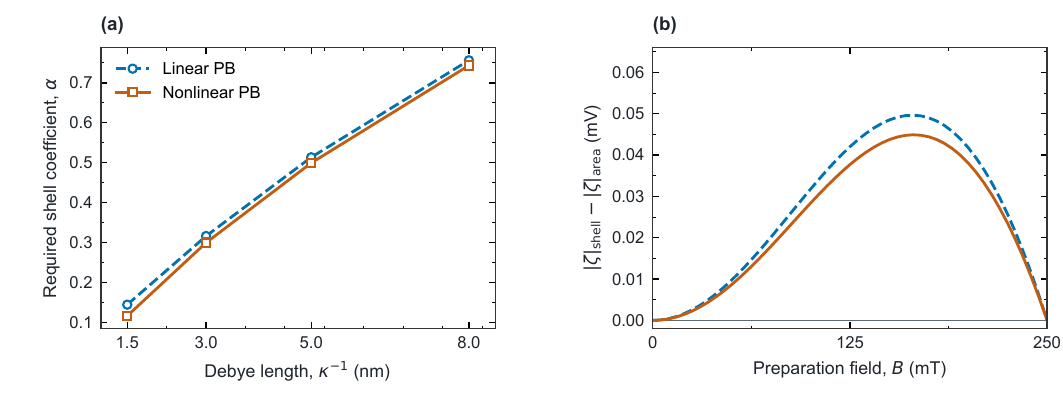}
\end{adjustwidth}
\caption{Comparison of linear and nonlinear electrostatics.
Blue dashed and orange solid curves denote the linear and nonlinear
Poisson--Boltzmann calculations, respectively.
(\textbf{a}) Shell coefficient determined from the two endpoint potentials
at \(\kappa^{-1}=1.5,3,5,8\,\mathrm{nm}\),
\(t_{\mathrm{lig}}=0\), and \(\delta_{\mathrm{slip}}=\SI{1}{nm}\).
Circles and squares mark the evaluated screening lengths; joining lines
guide the eye. (\textbf{b}) Difference of predicted magnitudes,
\(\Delta|\zeta|=|\zeta_{\mathrm{shell}}|-|\zeta_{\mathrm{area}}|\), at
\((\kappa^{-1},t_{\mathrm{lig}},\delta_{\mathrm{slip}})=(3,0,1)\,\mathrm{nm}\).
Each charge law is calibrated independently to the same endpoints with
quadratic field dependence. The nonlinear calculation uses a
\SI{5}{mT} grid and gives a maximum separation of \SI{0.0449}{mV}.
The curves compare model predictions, not experimental uncertainties.}
\label{fig:comparison}
\end{figure}

We next compare the linear result with the nonlinear calculation at
\((\kappa^{-1},t_{\mathrm{lig}},\delta_{\mathrm{slip}})=(3,0,1)\,\mathrm{nm}\).
The unamplified linear response is 0.575658, and matching the endpoint ratio
requires a shell coefficient \(\alpha=0.316017\). The nonlinear inverse
calculation gives \(\alpha_{\mathrm{NL}}=0.299307\), a decrease of 5.29\%
in this coefficient. The corresponding charge ratio decreases by 1.27\%,
from 0.075287 to 0.074331. At the observed endpoints, the nonlinear
charge-plane potential magnitudes \(|e\psi(R_s)|/(k_{\mathrm B}T)\) are
0.7568 and 0.6538. The nonlinear correction is therefore numerically
resolved, but it does not change the conditional shell interpretation
at this interface. Figure~\ref{fig:comparison}a shows how the correction
varies with screening length.

\begin{table}[htbp]
\centering
\caption{Endpoint magnitude ratios for
\((\kappa^{-1},t_{\mathrm{lig}},\delta_{\mathrm{slip}})=(3,0,1)\,\mathrm{nm}\).
For each electrostatic model, we independently calibrate the zero-field
charge to \(|\zeta_0|=\SI{13.2}{mV}\). The reported ratio provides an
endpoint constraint, with no assigned statistical uncertainty.}
\label{tab:benchmark}
\begin{adjustwidth}{-\extralength}{0cm}
\centering
\begin{tabular}{lcc}
\toprule
Charge closure & Linear PB & Nonlinear PB\\
\midrule
Constant total effective charge & 10.0625 & 4.83474\\
Constant charge-plane density & 0.628906 & 0.639167\\
Unamplified shell & 0.575658 & 0.585835\\
\midrule
Reported central ratio & \multicolumn{2}{c}{0.757576}\\
\bottomrule
\end{tabular}
\end{adjustwidth}
\end{table}

The constant-total-charge calculation shows why this electrostatic check
is needed. Linear theory gives \(|\zeta_B|=\SI{132.8}{mV}\), with a
back-propagated \(|e\psi(R_s)|/(k_{\mathrm B}T)=8.66\), which is far
outside the weak-potential regime. The nonlinear solution gives
\SI{63.8}{mV}, corresponding to a ratio of 4.83
(Table~\ref{tab:benchmark}). Across the four screening lengths evaluated,
the nonlinear constant-charge ratio ranges from 3.42 to 5.76. Although
nonlinearity changes the magnitude substantially, every tested scenario
still predicts an increase opposite to the reported trend. This conclusion
comes from the numerical solutions for these fixed-boundary scenarios;
we do not extend the linear inequality \(\rho_Q>1\) as a proof for the
nonlinear problem.

We calibrate the shell and area laws independently, using one
endpoint coefficient for each. The linear coefficients are
0.316017 and 0.204593, respectively, while the nonlinear values are
0.299307 and 0.189298. These differences in the fitted coefficients do
not produce comparably large differences in the predicted potentials:
the largest evaluated interior separations are only \SI{0.0496}{mV}
for the linear calculation and \SI{0.0449}{mV} for the nonlinear
calculation, near \SI{165}{mT} (Figure~\ref{fig:comparison}b).
Endpoint calibration therefore absorbs most of the finite-shell
correction. Distinguishing these fixed scenarios reliably would require
uncertainties well below their separation, and the present data do not
provide that precision assessment.

\Needspace{6\baselineskip}

\subsection{Exact identifiability and analytic parameter sensitivities}

The similarity of calibrated curves also raises the question of which
interfacial parameters the normalized response can determine. We make
the relevant combinations explicit by writing the screening factor in
terms of \(\kappa^{-1}\):
\begin{equation}
\rho(B)=[1+\alpha f(B)]\frac{n_{\mathrm{shell}}[x(B)]}{n_{\mathrm{shell}}(x_1)}
\frac{[r_1+t_{\mathrm{lig}}+\delta_{\mathrm{slip}}]
[r_1+t_{\mathrm{lig}}+\kappa^{-1}]}
{[r(B)+t_{\mathrm{lig}}+\delta_{\mathrm{slip}}]
[r(B)+t_{\mathrm{lig}}+\kappa^{-1}]}
\label{eq:composite}
\end{equation}
The three electrokinetic lengths appear only through the sums
\(t_{\mathrm{lig}}+\delta_{\mathrm{slip}}\) and
\(t_{\mathrm{lig}}+\kappa^{-1}\). We can therefore change the individual
lengths according to
\begin{equation}
t_{\mathrm{lig}}\mapsto t_{\mathrm{lig}}+c,\qquad
\delta_{\mathrm{slip}}\mapsto\delta_{\mathrm{slip}}-c,\qquad \kappa^{-1}\mapsto\kappa^{-1}-c
\label{eq:degeneracy}
\end{equation}
without changing the complete normalized shell-linked linear response.
Here, \(c\) is an arbitrary constant length for which the transformed
lengths remain admissible. For example,
\((\kappa^{-1},t_{\mathrm{lig}},\delta_{\mathrm{slip}})=(3,0,1)\) and
\((2.5,0.5,0.5)\,\mathrm{nm}\) give the same \(\alpha\) and the same
\(\rho(B)\) for any shared \(f(B)\). Additional field points can
constrain the response shape, but they cannot independently determine
all three lengths within this shell-linked linear formulation. Exchanging
the two sums also leaves the expression unchanged. This exact invariance
belongs to the stated linear model; we do not assert it for nonlinear
Poisson--Boltzmann electrostatics.

\begin{figure}[htbp]
\begin{adjustwidth}{-\extralength}{0cm}
\centering
\includegraphics[width=\linewidth]{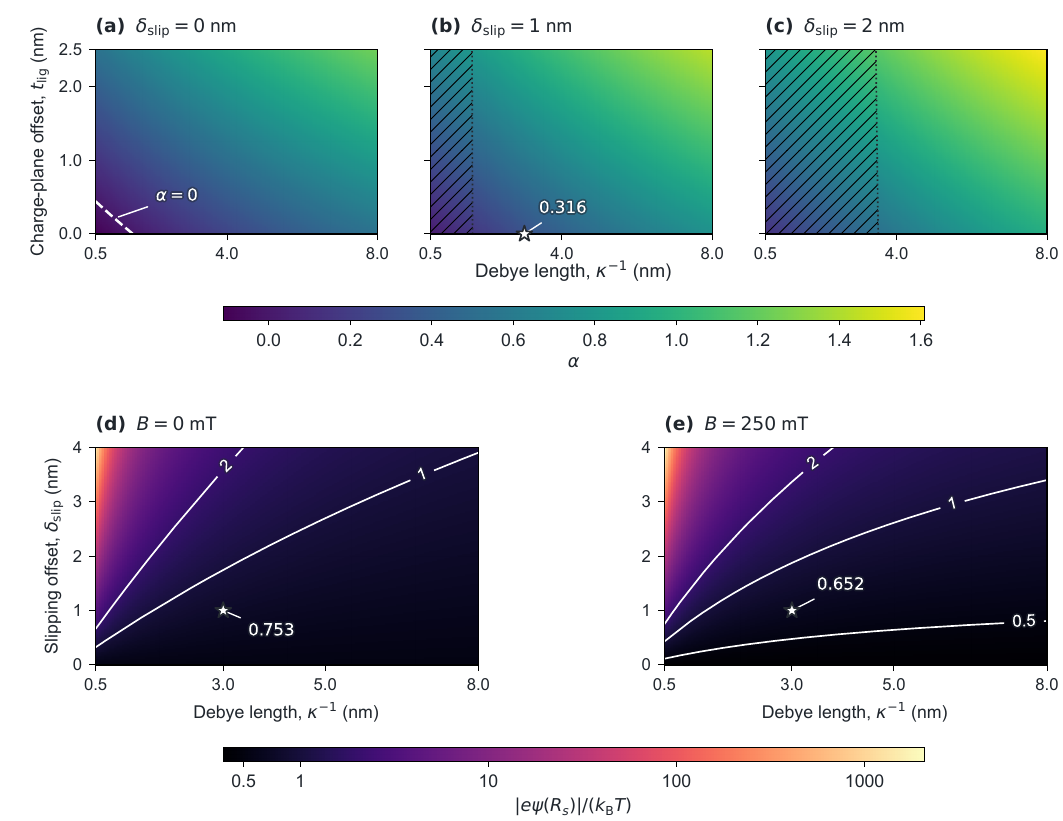}
\end{adjustwidth}
\caption{Effect of the charge-plane and slipping-plane offsets.
(\textbf{a--c}) Linear shell coefficient \(\alpha\) required by the
endpoint ratio, versus Debye length and charge-plane offset, for
\(\delta_{\mathrm{slip}}=0,1,2\,\mathrm{nm}\), respectively.
The panels share a color scale; dashed contours denote \(\alpha=0\)
where present. Hatching marks
\(|e\psi(R_s)|/(k_{\mathrm B}T)\geq1\) at either endpoint.
(\textbf{d,e}) Charge-plane potential magnitude in thermal units,
calculated from the linear solution at 0 and \SI{250}{mT}, respectively,
with \(t_{\mathrm{lig}}=0\). The color scale is logarithmic; white
contours identify values 0.5, 1, and 2 where they occur.
Stars identify the reference interface
\((\kappa^{-1},t_{\mathrm{lig}},\delta_{\mathrm{slip}})=(3,0,1)\,\mathrm{nm}\).
The number beside the star in (\textbf{b}) is \(\alpha=0.316\);
those in (\textbf{d,e}) are the thermal-potential ratios 0.753 and 0.652.
The unit threshold tests the weak-potential assumption; values below
unity do not by themselves establish quantitative accuracy.}
\label{fig:inference}
\end{figure}

To see how an assumed interface affects the inferred charge weighting,
we differentiate Equation~\eqref{eq:alpha_inverse} while holding the
observed ratio and structural inputs fixed:
\begin{align}
\frac{\partial\alpha}{\partial(\kappa^{-1})}
&=\frac{(1+\alpha)(r_1-r_2)}
{(r_2+t_{\mathrm{lig}}+\kappa^{-1})(r_1+t_{\mathrm{lig}}+\kappa^{-1})}
\nonumber\\
\frac{\partial\alpha}{\partial\delta_{\mathrm{slip}}}
&=\frac{(1+\alpha)(r_1-r_2)}
{(r_2+t_{\mathrm{lig}}+\delta_{\mathrm{slip}})(r_1+t_{\mathrm{lig}}+\delta_{\mathrm{slip}})}
\nonumber\\
\frac{\partial\alpha}{\partial t_{\mathrm{lig}}}
&=\frac{\partial\alpha}{\partial(\kappa^{-1})}
+\frac{\partial\alpha}{\partial\delta_{\mathrm{slip}}}
\label{eq:sensitivity}
\end{align}
All three derivatives are positive. Increasing one length while keeping
the other two fixed therefore requires greater charge weighting to
retain the observed ratio. At the reference interface, these derivatives are
0.107, 0.157, and \SI{0.264}{\per\nano\metre}, respectively.
Figures~\ref{fig:inference}a--c show the inferred coefficient as the
charge-plane offset and screening length vary at three fixed slipping
offsets, with a common full-range color scale. Changing
\(t_{\mathrm{lig}}\) changes both sums in Equation~\eqref{eq:composite},
whereas changing \(\delta_{\mathrm{slip}}\) alone changes only
\(t_{\mathrm{lig}}+\delta_{\mathrm{slip}}\). The value of the inferred
coefficient does not by itself establish the applicability of linear
electrostatics. We therefore vary the slipping offset explicitly in Figures~\ref{fig:inference}d,e to examine the weak-potential
assumption at each endpoint.

Although screening and the slipping offset both enter the electrostatic
response, they have different physical roles. The attenuation from the
charge plane to the slipping plane in linear spherical electrostatics is
\begin{equation}
\frac{\zeta}{\psi(R_s)}=
\frac{R_s}{R_s+\delta_{\mathrm{slip}}}\exp(-\kappa\delta_{\mathrm{slip}})
\label{eq:slip_attenuation}
\end{equation}
At a fixed charge-plane radius and potential, moving the slipping plane
outward by increasing \(\delta_{\mathrm{slip}}\) lowers the zeta magnitude.
For \(\delta_{\mathrm{slip}}>0\), increasing \(\kappa^{-1}\) weakens this
attenuation, whereas at \(\delta_{\mathrm{slip}}=0\) the attenuation
factor is unity for every \(\kappa^{-1}\). If the same
\(\delta_{\mathrm{slip}}\) and \(\kappa^{-1}\) apply at both field
conditions, the exponential factor cancels from the normalized endpoint
ratio, but the radius factors remain. The attenuation of the physical
potential and the charge coefficient inferred from an endpoint ratio
therefore describe different aspects of the problem. Neither offset map
directly measures an interfacial thickness.

Conversely, if the observed zeta magnitude is fixed, linear
back-propagation requires a larger charge-plane potential for a larger
slipping offset. A shorter Debye length has the same effect when
\(\delta_{\mathrm{slip}}>0\). The two endpoint maps identify where
this inferred potential reaches the thermal scale,
\(|e\psi(R_s)|/(k_{\mathrm B}T)=1\). They provide a diagnostic of the
weak-potential assumption, not confidence regions or measured limits
on \(\delta_{\mathrm{slip}}\); a value below unity alone does not
guarantee accuracy.

\subsection{What intermediate-field measurements can distinguish}

\begin{figure}[htbp]
\begin{adjustwidth}{-\extralength}{0cm}
\centering
\includegraphics[width=\linewidth]{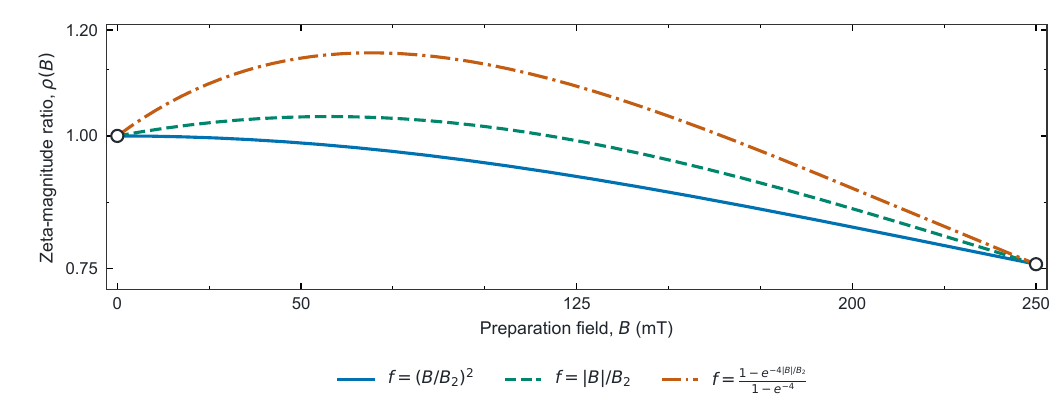}
\end{adjustwidth}
\caption{Normalized zeta magnitude for three charge-response functions:
quadratic \(f_2(B)=(B/B_2)^2\) (blue solid),
linear \(f_1(B)=|B|/B_2\) (green dashed), and
saturating \(f_{\mathrm{sat}}(B)=[1-e^{-4|B|/B_2}]/[1-e^{-4}]\)
(orange dash-dotted). The linear Poisson--Boltzmann calculations use
\((\kappa^{-1},t_{\mathrm{lig}},\delta_{\mathrm{slip}})=(3,0,1)\,\mathrm{nm}\)
and \(\alpha=0.316017\). Open circles mark the two experimental
endpoint constraints; the intermediate-field curves are predictions.}
\label{fig:response}
\end{figure}

We now keep the interface fixed and compare the selected field-response
functions. This produces a larger separation than the comparison between
the independently calibrated shell and area laws. At \SI{125}{mT}, the
quadratic, linear-in-\(|B|\), and early-saturating linear-PB examples give
\(\rho=0.9234\), 0.9910, and 1.0940, respectively
(Figure~\ref{fig:response}). For these interfacial parameters, even the closest
pair differs by \SI{0.892}{mV}. If we choose the field that maximizes
the minimum pairwise separation of these three fixed curves, we obtain
approximately \SI{109}{mT}, with a nearest-pair gap of \SI{0.910}{mV}.
The separation at \SI{125}{mT} lies within 2\% of this value. These
results indicate where the specified curves separate most clearly; their usefulness in an experiment also depends on measurement
precision and the validity of the assumed interface.

The interpretation depends on the interfacial assumptions. If those
parameters instead give \(\alpha=0\), all three response functions
produce exactly the same curve. We therefore need independent
constraints on screening and the interface before interpreting an
interior measurement as a test of \(f(B)\). Distinguishing selected
response shapes, comparing charge closures, and determining individual
electrokinetic lengths remain three different inference problems.

\FloatBarrier
\subsection{Size and potential distributions}

\begin{figure}[!htbp]
\begin{adjustwidth}{-\extralength}{0cm}
\centering
\includegraphics[width=\linewidth]{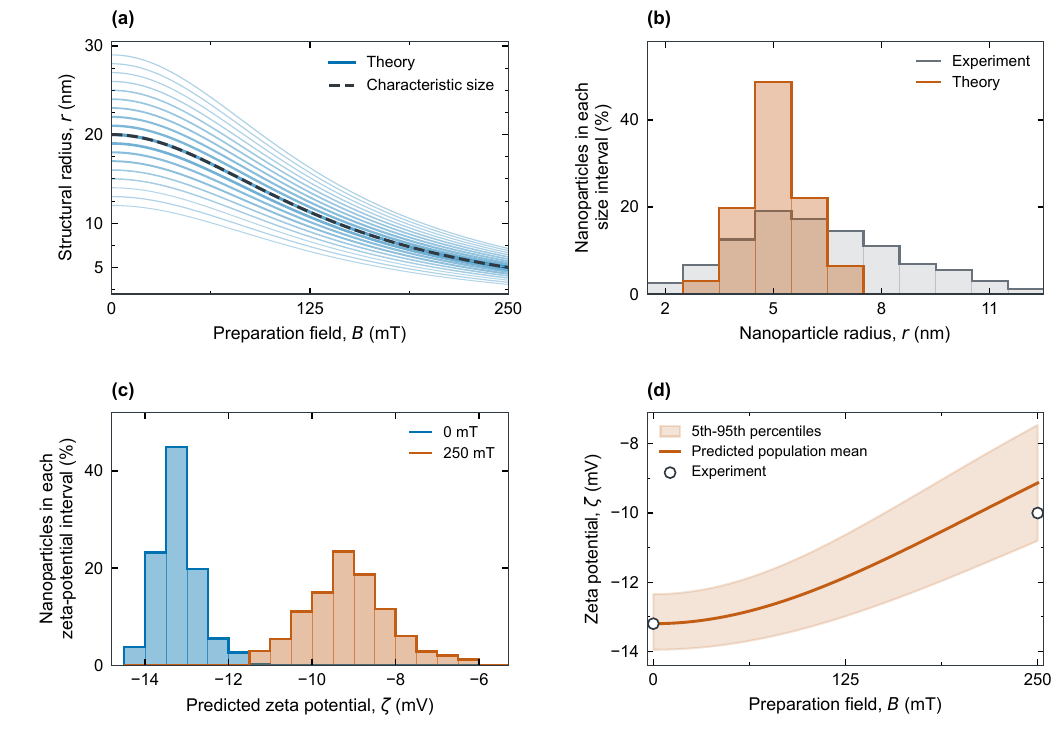}
\end{adjustwidth}
\caption{Nanoparticle size and zeta-potential distributions.
(\textbf{a}) Size-class trajectories from
Equation~\eqref{eq:population_flow}. Thicker, darker blue lines represent
larger initial particle fractions; the dark dashed curve is the
characteristic \SI{20}{nm}-to-\SI{5}{nm} trajectory, not a population mean.
(\textbf{b}) Percentages of nanoparticles in common \SI{1}{nm} radius
intervals at \SI{250}{mT}: grey for digitized experimental sizes and orange
for sizes predicted from the initial population.
(\textbf{c}) Percentages of nanoparticles in shared \SI{0.5}{mV}
zeta-potential intervals: blue at zero field and orange at \SI{250}{mT}.
Both potential histograms are calculated; each distribution in
(\textbf{b,c}) totals 100\%.
(\textbf{d}) Predicted number-weighted mean potential and shaded
5th--95th population percentiles. Open circles are the reported experimental
values, assumed to represent number-weighted means. The potential
\SI{-13.2}{mV} calibrates \(q_0\) over the measured initial sizes;
\SI{-10.0}{mV} calibrates \(\alpha_{\mathrm{pop}}=0.196860\)
over the measured endpoint sizes. Applying those charge parameters to the predicted endpoint
sizes gives \SI{-9.131}{mV}, which is the endpoint of the plotted line.
The calculations use
\((\kappa^{-1},t_{\mathrm{lig}},\delta_{\mathrm{slip}})=(3,0,1)\,\mathrm{nm}\)
and \(f(B)=(B/B_2)^2\). Initial size frequencies are preserved.
Moments and percentiles use the original size-bin centers; histogram
percentages use the within-bin construction in Section 2.4 and SI
Section S9.5. The shaded interval is population spread, not uncertainty
in the mean.}
\label{fig:population}
\end{figure}

Equation~\eqref{eq:population_flow} maps the 18 zero-field size classes
to the trajectories in Figure~\ref{fig:population}a. At \(B_2\),
the predicted centers span \(3.08675\)--\SI{7.14499}{nm}, with
mean \SI{5.09573}{nm} and standard deviation \SI{0.83515}{nm}.
The measured field-prepared population has mean \SI{6.270}{nm}
and standard deviation \SI{2.204}{nm}. Thus, although the structural
model reproduces its characteristic endpoint, its preserved-frequency
population extension gives a substantially narrower size distribution
(Figure~\ref{fig:population}b).

At \((\kappa^{-1},t_{\mathrm{lig}},\delta_{\mathrm{slip}})
=(3,0,1)\,\mathrm{nm}\), the zero-field potential distribution
has calibrated mean \SI{-13.2}{mV} and standard deviation
\SI{0.488}{mV}. With the measured-population charge calibration
\(\alpha_{\mathrm{pop}}=0.196860\), the structurally predicted
endpoint sizes give mean \SI{-9.13055}{mV} and standard deviation
\SI{0.99423}{mV} (Figure~\ref{fig:population}c,d).
Figure~\ref{fig:population}d therefore starts at the measured
zero-field mean but ends at \SI{-9.13055}{mV}, rather than
\SI{-10.0}{mV}. The latter value is reproduced by the same
\(\alpha_{\mathrm{pop}}=0.196860\) when the measured endpoint
sizes are used. The plotted curve instead averages over the structurally
predicted endpoint sizes. Its displacement is the consequence of this
change of size population, not failure to solve the charge-calibration
equations. SI Section S9.3 gives the exact ratio between the two endpoint
means.

The predicted endpoint relative size spread is 16.39\%, compared with
35.15\% for the measured size population. Its relative potential
spread is 10.89\%, compared with 21.73\% when the measured endpoint
sizes are evaluated with the same charge calibration.

{
The difference between the predicted and measured endpoint size distributions is further quantified by a one-dimensional Wasserstein distance of 1.414 nm (Supporting Information Section S9.3 and Figure S1a).}

Calibrating the charge coefficient to the structurally predicted
population instead gives \(\alpha_{\mathrm{pop}}=0.310829\),
an endpoint mean of \SI{-10.0}{mV}, and standard deviation
\SI{1.08890}{mV} (Figure S2). This endpoint constraint changes the
potential scale but not the predicted size distribution or relative
potential spread. The difference between predicted and measured size
widths therefore cannot be removed by a common charge rescaling.
SI Section S9.5 quantifies the small-size contribution from the
within-bin histogram representation.

\Needspace{6\baselineskip}
\section{Discussion}

The three-radius description connects the structural model to an
electrokinetic boundary problem while keeping the charge-bearing surface
distinct from the slipping plane. The shell count supplies a size
dependence that approaches area scaling for large particles and includes
the finite-shell correction at smaller radii. Under field-independent
interfacial conditions, this charge law gives a decreasing unamplified
zeta magnitude as the radius decreases. The result follows from the
competition between the declining shell population and the geometric
and screening factors in the spherical solution. It provides a
structural contribution to the response, while the coefficient
\(\alpha\) describes the additional change in effective charge per
shell count.

Nonlinear electrostatics changes the magnitude of the predicted response
by different amounts for the different charge laws. At the reference
interface, the inferred shell coefficient changes only modestly, whereas
the constant-charge potential is substantially reduced relative to the
linear result. Constant charge nevertheless predicts a trend opposite
to the reported decrease in all four nonlinear screening scenarios.
This conclusion applies to the specified fixed-boundary calculations;
changes in screening, permittivity, or boundary positions between
preparations could alter it. The nonlinear equation removes the
weak-potential approximation but still assumes sharp spherical
boundaries, uniform permittivity, and a symmetric electrolyte.

Relating effective charge to the shell population remains a constitutive
assumption. Adsorbed ions, surface oxidation, ligands, counterion
association, acid--base equilibria, and a biomolecular corona can all
contribute to interfacial charge \cite{Hunter1981,Delgado2005}.
The near coincidence of independently calibrated shell and area curves
shows that the present endpoints cannot identify a shell-specific
charging mechanism. The inferred coefficients differ, but their
predicted potentials are separated by less than \SI{0.05}{mV} at the
reference interface. Testing these charge laws therefore requires both
intermediate-field measurements and an uncertainty smaller than the
calculated separation.

Even an accurately measured normalized curve cannot determine all three
electrokinetic lengths independently within the linear shell model.
The continuous invariance of Equation~\eqref{eq:degeneracy} requires
additional information about screening or boundary positions, and
the exchange symmetry of the two length sums can leave further
ambiguity. Conductivity can reveal changes in the electrolyte, but
ionic composition is also required to estimate a Debye length. A
matched experimental series should therefore combine particle-size
distributions and electrophoretic mobility with pH, temperature,
conductivity, ionic composition, and surface or corona characterization.
Reporting the mobility-to-zeta conversion is particularly important
when comparing heterogeneous populations \cite{Delgado2005,Bhattacharjee2016}.
{
The population extension also retains the two-length-sum nonidentifiability of the linear electrokinetic model when the common charge amplitude is unconstrained (Supporting Information Section S9.2).}

The population calculation tests a different part of the model. The
initial size frequencies are carried through the structural relation
and then converted to potentials using a common charge law and common
interfacial parameters. The resulting endpoint size distribution is
narrower than measured, despite agreement at the characteristic
structural endpoint. With charge parameters determined from both
measured size populations, this difference also shifts the predicted
mean potential away from the reported value. Recalibrating the charge
coefficient restores that mean but leaves the size distribution and
relative potential width unchanged. Thus the discrepancy cannot be
removed by a common charge rescaling. It concerns the assumption of
preserved initial frequencies under a shared structural relation;
size-dependent formation, growth, or aggregation would require a
more general population description.

{
Recent cross-system analyses of protein-corona formation have identified nanoparticle size and zeta potential among the important predictors of protein adsorption \cite{10.1021/acsnano.5c08608}. These findings highlight the possible coupling between particle-size distributions and interfacial composition, beyond the structural size dependence explicitly considered in the present model.
The potential histograms are predictions under these population assumptions. Equating the reported zeta values with number-weighted mean potentials is not established by the electrophoretic measurement, and a common charge amplitude and interface may not describe every size class. 
}
The calculation also omits double-layer relaxation, surface conduction, nonspherical morphology, ion-specific effects, and soft-corona permeability \cite{Ohshima2006}. Digitization and finite size-bin width affect the population representation, while the smallest particles place additional demands on the continuum description; the sensitivity to the 2 nm endpoint bin is examined in Supporting Information Section S9.4. These limitations motivate measurements of the size and mobility distributions together, so that the structural response and the electrokinetic conversion can be assessed separately.

\Needspace{8\baselineskip}
\section{Conclusions}

We have coupled the sphere-packing description of nanoparticle size to
the potential at a distinct hydrodynamic slipping plane. For effective
charge proportional to the defective-shell population, decreasing radius
reduces the unamplified linear zeta magnitude at a fixed interface, while
the finite-shell correction quantifies the departure from surface-area
scaling. This identifies a structural contribution without assigning a
microscopic charging mechanism. Nonlinear electrostatics changes the reference shell coefficient
from 0.316 to 0.299, but constant total charge still predicts the opposite
trend to the reported decrease in all four evaluated nonlinear scenarios.
This result is conditional on the fixed screening and boundary positions
used in those calculations. Agreement with the endpoint potentials does
not, however, identify a shell-specific charging mechanism: independently
calibrated shell and area charge laws differ by less than \SI{0.05}{mV}
at intermediate fields for the reference interface. Furthermore, the
normalized linear response depends on the Debye length, charge-plane
offset, and slipping-plane offset through only two length sums, preventing
their independent determination even from a precisely measured curve.
The population calculation provides a separate test of the structural
extension. Propagating the initial size frequencies produces an endpoint
population narrower than measured. Charge calibration using the measured
size populations gives \SI{-9.131}{mV} on the predicted endpoint sizes,
whereas calibration using the predicted sizes imposes \SI{-10.0}{mV}.
This rescaling cannot correct the size-width discrepancy and leaves the
predicted relative zeta-potential width unchanged. Thus, the distributions
test the model beyond agreement at characteristic-size endpoints.
Together, these results identify the size-dependent contribution to zeta
potential and the limits of inferring interfacial properties from it.
Matched size distributions, electrophoretic measurements, and independent
characterization of screening and the particle interface are needed to
separate the structural and charging contributions.

\authorcontributions{Conceptualization, Project Administration and Supervision: M.F. Pereira.
Investigation, software and Visualization: Y. Tawalbeh.
Formal analysis, methodology,  validation, writing-original draft, writing-review and editing: Y. Tawalbeh and M.F. Pereira.
}

\funding{This research received no external funding.}

\dataavailability{The analysis uses
values reported in the cited publications. The accompanying source package
contains the digitized size frequencies, model inputs, numerical results,
and Python code for the electrostatic and population calculations.
It also includes the figure data and numerical verification tests.}

\institutionalreview{Not
applicable. This theoretical study uses published aggregate data.}

\informedconsent{Not applicable.}

\conflictsofinterest{The authors declare no conflict of interest.}

\section*{Supporting Information}

\setlength{\emergencystretch}{1em}
\setlength{\extralength}{0pt}

\setcounter{secnumdepth}{2}
\setcounter{section}{0}
\renewcommand{\thesection}{S\arabic{section}}

\renewcommand{\thesubsection}{\thesection.\arabic{subsection}}

\setcounter{equation}{0}
\renewcommand{\theequation}{S\arabic{equation}}

\setcounter{table}{0}
\renewcommand{\thetable}{S\arabic{table}}

\setcounter{figure}{0}
\renewcommand{\thefigure}{S\arabic{figure}}

\newcommand{\Rs}{R_s}
\newcommand{\dslip}{\delta_{\mathrm{slip}}}
\newcommand{\tlig}{t_{\mathrm{lig}}}
\newcommand{\Btwo}{B_2}
\newcommand{\nshell}{n_{\mathrm{shell}}}

\section{Inputs and Notation}

The calculation uses two reported central zeta potentials \cite{Kthiri2021},
\(\zeta_0=\SI{-13.2}{mV}\) and \(\zeta_B=\SI{-10.0}{mV}\), at
\(B=0\) and \(\Btwo=\SI{250}{mT}\), respectively. The structural inputs
from the preceding sphere-packing theory \cite{Tawalbeh2025} are
\(r_1=r(0)=\SI{20}{nm}\) and \(r_2=r(\Btwo)=\SI{5}{nm}\); both are fixed
throughout. Here \(x_1=r_1/a\) and \(x_2=r_2/a\). The zeta-magnitude ratio is
\begin{equation}
\rho_{\mathrm{obs}}=\frac{|\zeta_B|}{|\zeta_0|}
=\frac{10.0}{13.2}=0.757576\ldots 
\label{eqS:rhoobs}
\end{equation}
Calculations retain full precision; inferred values are generally reported
to three significant figures. The interfacial parameters are specified model
conditions rather than measured quantities. Sections S1--S8 describe the
characteristic-particle calculation, and Section S9 extends it to the
measured size distributions.

The structural variables are \(r,x,n(x)\). The effective charge-plane radius
is \(\Rs=r+\tlig\), and the slipping plane is at \(\Rs+\dslip\).
The displacement \(\tlig\) locates the effective charge plane and does not
imply a measured ligand thickness. The inverse screening length is \(\kappa\), so the Debye length is
\(\kappa^{-1}\). The dimensionless structural shell depth \(\delta\) and
the physical slipping-plane displacement \(\dslip\) are distinct.
The field \(B\) refers to the condition under which the particles were prepared.

\begin{table}[htbp]
\centering
\caption{Model symbols and their status.}
\begin{adjustwidth}{-\extralength}{0cm}
\centering
\begin{tabular}{lll}
\toprule
Symbol & Meaning & Status \\
\midrule
\(B\) & Preparation-field condition & reported endpoint condition \\
\(r\) & Structural particle radius & theoretical input/trajectory \\
\(x=r/a\) & Dimensionless structural radius & inherited notation \\
\(n(x)\) & Effective total atom count & inherited packing model \\
\(\nshell(x)\) & Effective shell atom count & inherited shell contribution \\
\(\Rs=r+\tlig\) & Effective charge-plane radius & scenario-dependent boundary \\
\(\Rs+\dslip\) & Hydrodynamic slipping-plane radius & scenario-dependent boundary \\
\(\tlig\) & Charge-plane displacement & scenario parameter \\
\(\dslip\) & Slipping-plane displacement & scenario parameter \\
\(\kappa^{-1}\) & Debye length & scenario parameter \\
\(q_0\) & Zero-field effective charge weight & calibrated amplitude \\
\(\alpha\) & Endpoint charge-weighting response & inverse-model output \\
\bottomrule
\end{tabular}
\end{adjustwidth}
\end{table}

\section{Sphere-Packing Decomposition and Structural Trajectory}

The structural description combines classical nucleation theory with
sphere packing \cite{Kashchiev2000,Tawalbeh2025}. Its application to silver particles under
parallel and perpendicular magnetic-field configurations was examined in
Ref.~\cite{Tawalbeh2026APB}. A subsequent extension includes permanent-moment
alignment and recovers the silver-particle relation as the susceptibility-only
limit \cite{Tawalbeh2026}. We use this limit here.

Let \(x=r/a\), where \(a=\SI{0.144}{nm}\), and let the dimensionless shell
depth be \(\delta=4\) \cite{Tawalbeh2025}. All structural and population
radii considered here obey \(x>\delta\). The effective atom count is
\begin{equation}
n(x)=\phi_{\mathrm b}(x-\delta)^3
+\phi_{\mathrm d}\left[x^3-(x-\delta)^3\right]
\label{eqS:n}
\end{equation}
with \(\phi_{\mathrm b}=\pi/(3\sqrt{2})\), the congruent-sphere packing ceiling
\cite{Hales2005}, and \(\phi_{\mathrm d}=0.5172879\), the defective-shell
packing fraction determined in Ref.~\cite{Tawalbeh2025}. Its shell contribution is
\begin{equation}
\nshell(x)=\phi_{\mathrm d}\left[x^3-(x-\delta)^3\right]
=\phi_{\mathrm d}\left(3\delta x^2-3\delta^2x+\delta^3\right)
\label{eqS:shell}
\end{equation}
These continuum expressions represent the total and shell atom counts.
The geometric and packing-weighted shell fractions used in the main
article are distinct:
\begin{align}
\text{geometric shell fraction}
&=\frac{x^3-(x-\delta)^3}{x^3}
\nonumber\\
\text{packing-weighted shell fraction}
&=\frac{\nshell(x)}{n(x)}
\label{eqS:shell_fractions}
\end{align}
The first is a volume fraction and does not include packing efficiency.
The second is the fraction of model atoms assigned to the defective
shell. Between the characteristic radii of \SI{20}{nm} and \SI{5}{nm},
these fractions change from 0.0839 to 0.307 and from 0.0602 to 0.237,
respectively. Both fractions increase even though the absolute shell
count decreases.
The proportionality between effective charge and \(\nshell\) is a
constitutive assumption; the charge weight need not equal the microscopic
charge of an individual shell atom.
Neither the structural shell depth nor its physical thickness \(a\delta\)
is identified with an electrostatic screening length.

The structural trajectory follows from the spherical work of formation.
The orientation factor is fixed to unity, corresponding to deterministic
orientation:
\begin{equation}
\Delta F(x,B)
=-n(x)\left[\Delta\mu+E_B\left(\frac{B}{\Btwo}\right)^2\right]+E_sx^2,
\qquad E_s=4\pi\gamma a^2
\label{eqS:freeenergy}
\end{equation}
Here \(\Delta\mu\) is the zero-field chemical driving energy per atom,
\(\gamma\) is the interfacial energy per area, and \(E_B\) is the
field-associated driving energy per atom at \(\Btwo\). For a spherical
diamagnetic particle \cite{Tawalbeh2025},
\begin{equation}
E_B=\frac{V_a}{2\mu_0}\frac{3|\chi_m|}{3+\chi_m}B_2^2,
\qquad V_a=\frac{4\pi a^3}{3}
\label{eqS:magnetic_energy}
\end{equation}
Here \(\chi_m\) is the volume magnetic susceptibility, \(\mu_0\) is the
vacuum permeability, and \(V_a\) is the building-unit volume in this
structural convention. The field is expressed in consistent SI units when
evaluating this energy. These terms describe nucleation rather than
electrostatic charging. Stationarity gives
\begin{equation}
\frac{x}{n'(x)}
=\frac{\Delta\mu+E_B(B/\Btwo)^2}{2E_s},
\qquad
n'(x)=3\phi_{\mathrm b}x^2
+6\delta(\phi_{\mathrm d}-\phi_{\mathrm b})x
+3\delta^2(\phi_{\mathrm b}-\phi_{\mathrm d})
\label{eqS:nderivative}
\end{equation}
The stationarity condition follows by setting
\(\partial\Delta F/\partial x
=-n'(x)[\Delta\mu+E_B(B/\Btwo)^2]+2E_sx=0\).
It identifies a barrier maximum on the physical branch. Indeed,
\begin{equation}
\left.\frac{\partial^2\Delta F}{\partial x^2}\right|_{\partial_x\Delta F=0}
=2E_s\left[1-\frac{x n''(x)}{n'(x)}\right]<0
\qquad
n''(x)=6\phi_{\mathrm b}x+6\delta(\phi_{\mathrm d}-\phi_{\mathrm b})
\label{eqS:barrier_curvature}
\end{equation}
For \(x>\delta\), the numerator
\(n'(x)-xn''(x)
=3\delta^2(\phi_{\mathrm b}-\phi_{\mathrm d})-3\phi_{\mathrm b}x^2\)
is negative, whereas \(n'(x)\) and \(E_s\) are positive.
Thus the selected stationary size separates shrinkage of subcritical
clusters from post-critical growth; it is not a stable minimum of the work
of formation. The plotted structural trajectory retains the endpoint
identification made in the preceding size model and does not independently
describe the kinetics of subsequent growth.

Figure 1b of the main article illustrates this competition in arbitrary
size and energy units. Its two schematic curves are \(3u^2-2c u^3\), where
\(u\) is a dimensionless schematic coordinate and \(c=1\) and 1.45 distinguish
the two schematic driving strengths. Differentiation gives maxima at
\(u=1/c\), of height \(1/c^2\); the second derivative there is \(-6\).
The dots and vertical guides locate these maxima. The origin denotes the
parent phase and the descending branches indicate post-critical growth.
These schematic parameters are not inferred formation energies or additional
parameters of Equation~\eqref{eqS:freeenergy}.

Eliminating the two driving-energy combinations using the endpoint values
gives
\begin{equation}
\frac{x(B)}{n'[x(B)]}
=\frac{x_1}{n'(x_1)}
+\left[\frac{x_2}{n'(x_2)}-\frac{x_1}{n'(x_1)}\right]
\left(\frac{B}{\Btwo}\right)^2,
\qquad r(B)=a x(B)
\label{eqS:trajectory}
\end{equation}
The work of formation and the two endpoint conditions therefore determine
the intermediate trajectory. Equivalently, \(x(B)\) solves
\begin{align}
0={}&3\phi_{\mathrm b}x(B)^2+3\delta^2(\phi_{\mathrm b}-\phi_{\mathrm d})
\nonumber\\
&+\left\{6\delta(\phi_{\mathrm d}-\phi_{\mathrm b})
-\left[
\frac{x_1}{n'(x_1)}
+\left(\frac{x_2}{n'(x_2)}-\frac{x_1}{n'(x_1)}\right)
\left(\frac{B}{\Btwo}\right)^2
\right]^{-1}\right\}x(B)
\label{eqS:xbranch}
\end{align}
The larger root with \(x>\delta\), continuous between the specified
endpoints, is used. The endpoint sizes are imposed, while the intermediate
sizes are calculated from the model. At the endpoints,
\begin{equation}
x_1=138.8889,\qquad x_2=34.7222,\qquad
\frac{\nshell(x_2)}{\nshell(x_1)}=0.0572083
\label{eqS:shellratio}
\end{equation}
The common shell packing fraction cancels in this ratio.

\section{Linear Spherical Poisson--Boltzmann Derivation}

To distinguish position in the liquid from the structural particle radius
\(r\), let \(R\) denote the radial coordinate measured from the particle
center. For \(R\geq\Rs\), the assumed weak-potential, uniform-permittivity
model is the linear spherical Poisson--Boltzmann equation
\cite{Hunter1981,Ohshima2006}:
\begin{equation}
\frac{1}{R^2}\frac{d}{dR}\left(R^2\frac{d\psi}{dR}\right)
=\kappa^2\psi
\label{eqS:PB}
\end{equation}
The effective total charge \(Q_{\mathrm{eff}}\) is represented on the sharp
plane \(R=\Rs\), with liquid-side boundary condition
\begin{equation}
-\varepsilon\left.\frac{d\psi}{dR}\right|_{\Rs^+}
=\frac{Q_{\mathrm{eff}}}{4\pi\Rs^2}
=\sigma_{\mathrm{eff}},
\qquad \psi(\infty)=0
\label{eqS:boundary}
\end{equation}
Here \(\varepsilon\) is the absolute liquid permittivity and
\(\sigma_{\mathrm{eff}}\) is the same effective charge expressed per
charge-plane area, not an independent fitted charge.
Multiplication of Equation~\eqref{eqS:PB} by \(R\) gives
\(d^2(R\psi)/dR^2=\kappa^2(R\psi)\). The condition at infinity removes
the exponentially increasing solution. The remaining solution and its
derivative at the charge plane satisfy
\begin{align}
\psi(R)&=\psi(\Rs)\frac{\Rs}{R}\exp[-\kappa(R-\Rs)]
\nonumber\\
-\psi'(\Rs)&=\left(\kappa+\frac{1}{\Rs}\right)\psi(\Rs)
\qquad
\psi(\Rs)=\frac{Q_{\mathrm{eff}}}
{4\pi\varepsilon\Rs(1+\kappa\Rs)}
\label{eqS:PB_solution_steps}
\end{align}
Inserting this charge-plane value into the decaying solution gives
\begin{equation}
\psi(R)=\frac{Q_{\mathrm{eff}}}{4\pi\varepsilon R(1+\kappa\Rs)}
\exp[-\kappa(R-\Rs)]
\label{eqS:psi}
\end{equation}
At the slipping plane,
\begin{equation}
\zeta=\psi(\Rs+\dslip)
=\frac{Q_{\mathrm{eff}}}
{4\pi\varepsilon(\Rs+\dslip)(1+\kappa\Rs)}
\exp(-\kappa\dslip)
\label{eqS:zeta}
\end{equation}
The attenuation between the two effective planes is
\begin{equation}
\frac{\zeta}{\psi(\Rs)}
=\frac{\Rs}{\Rs+\dslip}\exp(-\kappa\dslip)
\label{eqS:attenuation}
\end{equation}
Equations~\eqref{eqS:psi}--\eqref{eqS:attenuation} do not include
electrophoretic relaxation, surface conduction, nonuniform permittivity,
ion-specific interactions, or soft-particle permeability.

\section{Normalized Response and Endpoint Inversion}

Applying Equation~\eqref{eqS:zeta} at \(B\) and at zero field gives
\begin{align}
\rho(B)&=\frac{|\zeta(B)|}{|\zeta(0)|}\nonumber\\
&=\left|\frac{Q_{\mathrm{eff}}(B)}{Q_{\mathrm{eff}}(0)}\right|
\frac{\varepsilon(0)}{\varepsilon(B)}
\frac{\Rs(0)+\dslip(0)}{\Rs(B)+\dslip(B)}
\frac{1+\kappa(0)\Rs(0)}{1+\kappa(B)\Rs(B)}
\nonumber\\
&\quad{}\times
\exp[\kappa(0)\dslip(0)-\kappa(B)\dslip(B)]
\label{eqS:ratio}
\end{align}
For the shell-linked charge hypothesis,
\begin{equation}
Q_{\mathrm{eff}}(B)
=q_0[1+\alpha f(B)]\nshell[x(B)],
\qquad f(0)=0,\quad f(\Btwo)=1
\label{eqS:closure}
\end{equation}
where \(q_0\) is field invariant and \(1+\alpha f(B)>0\).
The shell packing fraction remains fixed: \(\alpha\) changes the effective
charge weight, not the structural packing. We use
\(f(B)=(B/\Btwo)^2\). Under field-invariant permittivity, screening, and
offsets, the common \(q_0\) and the exponential attenuation cancel:
\begin{align}
\left.\rho(B)\right|_{\alpha=0}
&=\frac{\nshell[x(B)]}{\nshell(x_1)}
\frac{r_1+\tlig+\dslip}{r(B)+\tlig+\dslip}
\frac{1+\kappa(r_1+\tlig)}{1+\kappa[r(B)+\tlig]}
\nonumber\\
\rho(B)&=[1+\alpha f(B)]\left.\rho(B)\right|_{\alpha=0},
\qquad
\alpha
=\frac{\rho_{\mathrm{obs}}}{\left.\rho(\Btwo)\right|_{\alpha=0}}-1
\label{eqS:T_invariant}
\end{align}
The response \(\left.\rho(B)\right|_{\alpha=0}\) is evaluated with a
field-independent charge weight. For
\((\kappa^{-1},\tlig,\dslip)=(3,0,1)\,\mathrm{nm}\),
\begin{equation}
\left.\rho(\Btwo)\right|_{\alpha=0}
=0.0572083\left(\frac{21}{6}\right)
\left(\frac{1+20/3}{1+5/3}\right)=0.575658,
\qquad \alpha=0.316017
\label{eqS:Tworked}
\end{equation}
The value \(\alpha=0.316017\) therefore means that the effective charge
per shell atom at \(\Btwo\) is \(1.316017\,q_0\), while the total charge
also changes through \(\nshell(x_2)/\nshell(x_1)\). It does not mean that
the total particle charge increases by 31.6\%.
For the negative potentials considered here \(q_0<0\); a positive
\(\alpha\) increases the magnitude of this effective charge weight.
Because the endpoint potential determines the charge coefficient, its
reproduction does not independently test the shell hypothesis. The
coefficient is conditional on the specified screening and boundary
positions, not an independently measured material constant.

\Needspace{10\baselineskip}
\section{Analytic Response Properties and Identifiability}

\subsection{Finite-shell correction and monotonicity}

Factoring the shell expression gives
\begin{equation}
\nshell(r/a)=\frac{3\phi_{\mathrm d}\delta r^2}{a^2}
\left[1-\frac{a\delta}{r}+\frac{(a\delta)^2}{3r^2}\right],
\qquad r>a\delta=\SI{0.576}{nm}
\label{eqS:finite}
\end{equation}
The derivative of the expression in square brackets is
\(a\delta/r^2-2(a\delta)^2/(3r^3)>0\). Consequently,
\begin{align}
\frac{d\ln\!\left[\left.\rho(B)\right|_{\alpha=0}\right]}{dr}
={}&\frac{2}{r}
+\frac{a\delta/r^2-2(a\delta)^2/(3r^3)}
{1-a\delta/r+(a\delta)^2/(3r^2)}
\nonumber\\
&-\frac{1}{r+\tlig+\dslip}
-\frac{1}{r+\tlig+\kappa^{-1}}>0
\label{eqS:monotonic}
\end{align}
Here differentiation is along the radius-dependent response in
Equation~\eqref{eqS:T_invariant}, with interface parameters fixed.
Both subtracted terms are at most \(1/r\), while the shell correction
contributes positively. This proves that the unamplified linear shell response
decreases along a decreasing-radius trajectory for every field-invariant
nonnegative offset pair and positive Debye length.

Each radial factor in the endpoint response lies between one and
\(r_1/r_2=4\), giving
\begin{equation}
\frac{\nshell(x_2)}{\nshell(x_1)}
<\left.\rho(\Btwo)\right|_{\alpha=0}
<\frac{\nshell(x_2)}{\nshell(x_1)}
\left(\frac{r_1}{r_2}\right)^2
=0.915332
\label{eqS:bounds}
\end{equation}
The bounds are approached as both additive length combinations
\(\tlig+\dslip\) and \(\tlig+\kappa^{-1}\) become large or tend to zero,
respectively. The upper bound is less than one because the finite-shell
correction in square brackets in Equation~\eqref{eqS:finite} is smaller
at \(r_2\) than at \(r_1\). This bounds the unamplified response, not
the charge coefficient \(\alpha\).

\subsection{Exact composite-parameter degeneracy}

Written directly in the physical lengths, the same response is
\begin{equation}
\left.\rho(B)\right|_{\alpha=0}
=\frac{\nshell[r(B)/a]}{\nshell(r_1/a)}
\frac{(r_1+\tlig+\dslip)(r_1+\tlig+\kappa^{-1})}
{[r(B)+\tlig+\dslip][r(B)+\tlig+\kappa^{-1}]}
\label{eqS:composite}
\end{equation}
The transformation
\[
(\tlig,\dslip,\kappa^{-1})\mapsto
(\tlig+c,\dslip-c,\kappa^{-1}-c)
\]
leaves both length sums unchanged wherever the transformed lengths remain
admissible; \(c\) is a displacement used only to express this invariance.
It therefore preserves the unamplified normalized response, the endpoint
coefficient, and every normalized curve with a shared \(f(B)\).
The two triples
\((\kappa^{-1},\tlig,\dslip)=(3,0,1)\) and
\((2.5,0.5,0.5)\,\mathrm{nm}\) are an explicit example.
Interchanging the two additive length sums also preserves the expression.
Thus the constituent lengths are structurally nonidentifiable in the
normalized shell-linked linear map, even if the entire field curve were
measured without noise.

When the overall charge scale remains free, the absolute linear response is
\begin{equation}
|\zeta(B)|
=\frac{|q_0|\kappa^{-1}\exp(-\kappa\dslip)}{4\pi\varepsilon}
\frac{[1+\alpha f(B)]\nshell[r(B)/a]}
{[r(B)+\tlig+\dslip][r(B)+\tlig+\kappa^{-1}]}
\label{eqS:absolute_degeneracy}
\end{equation}
Changing \(q_0\) by the inverse change of
\(\kappa^{-1}\exp(-\kappa\dslip)\) preserves the absolute response as well.
Independent screening or boundary information can remove parts of this
ambiguity. Nonlinear Poisson--Boltzmann solutions are not assumed to possess
this exact invariance.

\subsection{Inverse sensitivities and the zero-response boundary}

Solving the endpoint condition for the charge coefficient gives
\begin{equation}
1+\alpha
=\rho_{\mathrm{obs}}\frac{\nshell(x_1)}{\nshell(x_2)}
\frac{r_2+\tlig+\dslip}{r_1+\tlig+\dslip}
\frac{r_2+\tlig+\kappa^{-1}}{r_1+\tlig+\kappa^{-1}}
\label{eqS:Ainverse}
\end{equation}
Logarithmic differentiation gives
\begin{align}
\frac{\partial\alpha}{\partial(\kappa^{-1})}
&=(1+\alpha)
\frac{r_1-r_2}{(r_2+\tlig+\kappa^{-1})(r_1+\tlig+\kappa^{-1})}
\nonumber\\
\frac{\partial\alpha}{\partial\dslip}
&=(1+\alpha)
\frac{r_1-r_2}{(r_2+\tlig+\dslip)(r_1+\tlig+\dslip)}
\nonumber\\
\frac{\partial\alpha}{\partial\tlig}
&=\frac{\partial\alpha}{\partial\dslip}
+\frac{\partial\alpha}{\partial(\kappa^{-1})}
\label{eqS:derivatives}
\end{align}
The worked values, in the order displayed, are 0.107284, 0.156669, and
\SI{0.263953}{\per\nano\metre}. The derivatives were also checked by
central finite differences. Their positive signs show that a longer
Debye length, a more distant slipping plane, or a more distant charge
plane requires a larger charge-weight multiplier to reproduce the same
two measured potentials. Screening and the slipping-plane offset enter
as separate lengths but cannot be separated completely from a normalized
response because of Equation~\eqref{eqS:composite}.

For fixed \(\tlig,\dslip\), the zero-response boundary is the positive
screening length satisfying
\begin{equation}
\rho_{\mathrm{obs}}\frac{\nshell(x_1)}{\nshell(x_2)}
\frac{r_2+\tlig+\dslip}{r_1+\tlig+\dslip}
\frac{r_2+\tlig+\kappa^{-1}}{r_1+\tlig+\kappa^{-1}}=1
\label{eqS:zerocontour}
\end{equation}
A nonpositive solution does not describe an admissible finite-screening
case. At zero offsets the shell threshold is \SI{1.491807}{nm}; at
\(\tlig=0,\dslip=\SI{1}{nm}\), it is \SI{0.388805}{nm}, below the
plotted inverse-map domain. Thus the \(\dslip=\SI{1}{nm}\) inverse-map
slice has no zero contour within its displayed range. The
\(\dslip=0\) slice does include a zero contour, consistent with the
zero-offset threshold above. For the constant-area closure at zero offsets,
replacing \(\nshell(x_2)/\nshell(x_1)\) by \((r_2/r_1)^2\) gives
\SI{2.388060}{nm}.

\section{Response Shapes and Equal-Flexibility Comparisons}

The quadratic and two alternative charge-response functions are
\begin{equation}
f_2(B)=\left(\frac{B}{\Btwo}\right)^2,\qquad
f_1(B)=\frac{|B|}{\Btwo},\qquad
f_{\mathrm{sat}}(B)
=\frac{1-\exp(-4|B|/\Btwo)}{1-\exp(-4)}
\label{eqS:shapes}
\end{equation}
These functions vary the assumed field dependence while preserving
\(f(0)=0\) and \(f(\Btwo)=1\). The quadratic function is the lowest nonzero
analytic scalar response under field-reversal symmetry. The other two
functions are nonanalytic in signed field at zero and are used as
endpoint-preserving shape examples. The saturation constant 4 is specified
rather than fitted.

With a fixed reference interface and positive \(\alpha\), the
linear-versus-quadratic ratio difference is
\begin{equation}
\rho_1(B)-\rho_2(B)
=\alpha\left.\rho(B)\right|_{\alpha=0}
\frac{|B|}{\Btwo}\left(1-\frac{|B|}{\Btwo}\right)
\label{eqS:shape_gap}
\end{equation}
Here subscripts 1 and 2 label the response shapes, not the two structural
endpoint indices. Numerically maximizing the minimum pairwise gap among the
three reference curves gives \(B=\SI{108.75}{mT}\), with a closest-pair
separation of 0.068943 in ratio, or \SI{0.91004}{mV}. At \SI{125}{mT}
the closest gap is \SI{0.89247}{mV}. These are separation measures for fixed
scenarios under an equal-noise criterion; no experimental error distribution
or statistical power is inferred. If \(\alpha=0\), all response shapes
coincide and this discrimination disappears.

To compare shell and area closures fairly, each receives an independent
coefficient while the same field law, structural trajectory, and interface
are held fixed. The area closure replaces \(\nshell[x(B)]/\nshell(x_1)\)
by \(\Rs(B)^2/\Rs(0)^2\) in the normalized charge response.
For \(\tlig=0\), the unamplified responses obey
\begin{equation}
\frac{\left.\rho_{\mathrm{shell}}(B)\right|_{\alpha=0}}
{\left.\rho_{\mathrm{area}}(B)\right|_{\alpha=0}}
=\frac{1-a\delta/r(B)+(a\delta)^2/[3r(B)^2]}
{1-a\delta/r_1+(a\delta)^2/(3r_1^2)}
\label{eqS:fairratio}
\end{equation}
At the reference interface the linear coefficients are 0.316017 and
0.204593. Their optimized maximum interior zeta-magnitude separation is
\SI{0.049640}{mV} at \SI{164.58}{mT}. The nonlinear comparison in
Section S7 separately recalibrates both coefficients; its largest separation
on the \SI{5}{mT} grid is \SI{0.044895}{mV} at \SI{165}{mT}.
The endpoint constraints absorb much of the finite-shell correction, so
the different coefficients should not be interpreted as an easily measurable
difference between mechanisms.

\section{Nonlinear Electrostatics and Solver Verification}

\subsection{Boundary problem and charge inversion}

For a symmetric monovalent electrolyte, the signed potential in the same
physical radial coordinate \(R\) obeys
\begin{equation}
\frac{d^2\psi}{dR^2}+\frac{2}{R}\frac{d\psi}{dR}
=\kappa^2\frac{k_{\mathrm B}T}{e}
\sinh\!\left(\frac{e\psi}{k_{\mathrm B}T}\right),
\qquad R\geq\Rs
\label{eqS:NPB}
\end{equation}
Here \(k_{\mathrm B}T/e=\SI{25.6926}{mV}\) at
\(T=\SI{298.15}{K}\), where \(e\) is the positive elementary charge
and \(k_{\mathrm B}\) is Boltzmann's constant. For this electrolyte,
the inverse Debye length is related to the bulk ion number density by
\begin{equation}
\kappa^2=\frac{2e^2c_\infty}{\varepsilon k_{\mathrm B}T}
\label{eqS:Debye_length}
\end{equation}
Here \(c_\infty\) is the bulk number density of either ionic species;
it is not a molar concentration or a nanoparticle atom count.
The factor of two includes the equally abundant positive and negative
monovalent ions. The present calculations specify \(\kappa^{-1}\)
as an interfacial condition rather than infer a measured salt
concentration. Products such as \(\kappa R\), \(\kappa\Rs\), and
\(\kappa\dslip\) are dimensionless, with the radii and screening
length expressed in the same units.
The boundary conditions retain the same effective
charge and slipping-plane definitions:
\begin{equation}
-\varepsilon\psi'(\Rs)
=\frac{Q_{\mathrm{eff}}}{4\pi\Rs^2},
\qquad \psi(\infty)=0,\qquad
\zeta=\psi(\Rs+\dslip)
\label{eqS:NPB_boundary}
\end{equation}
The signed boundary conditions give a negative potential for negative charge.
The nonlinear equation reduces to Equation~\eqref{eqS:PB} when
\(|e\psi/(k_{\mathrm B}T)|\ll1\), because
\(\sinh[e\psi/(k_{\mathrm B}T)]\) then approaches
\(e\psi/(k_{\mathrm B}T)\). This is the approximation tested by the
charge-plane diagnostic below; a small slipping-plane potential alone
need not guarantee a small potential closer to the charge plane.

At the finite remote boundary \(R_{\max}\), the decaying weak-field spherical
tail is imposed:
\begin{equation}
\psi'(R_{\max})
+\left(\kappa+\frac{1}{R_{\max}}\right)\psi(R_{\max})=0
\label{eqS:tail}
\end{equation}
The numerical coordinate \(R_{\max}\) specifies truncation of the liquid
domain, not a particle boundary. This condition approximates the far field
where the solved potential is exponentially small, rather than linearizing
the near-field equation.

For inverse charge recovery, the exterior problem starts at
\(R=\Rs+\dslip\) with \(\psi(\Rs+\dslip)=\zeta\), followed by inward
integration to \(\Rs\). The resulting derivative supplies
\(Q_{\mathrm{eff}}=-4\pi\varepsilon\Rs^2\psi'(\Rs)\).
A separate forward charge-boundary solve verifies that this value returns
the prescribed slipping-plane potential. For field-invariant permittivity
and temperature, the ratio of
\(|Q_{\mathrm{eff}}|/[4\pi\varepsilon(k_{\mathrm B}T/e)]\)
equals the absolute-charge ratio, so no numerical permittivity is needed
for that ratio. The shell inversion is
\begin{equation}
\alpha_{\mathrm{NL}}
=\frac{|Q_{\mathrm{eff}}(\Btwo)|}{|Q_{\mathrm{eff}}(0)|}
\frac{\nshell(x_1)}{\nshell(x_2)}-1
\label{eqS:alphaNL}
\end{equation}
The calibrated nonlinear field law uses
\[
Q_{\mathrm{eff}}(B)
=Q_{\mathrm{eff}}(0)
\frac{\nshell[x(B)]}{\nshell(x_1)}
[1+\alpha_{\mathrm{NL}}(B/\Btwo)^2]
\]
For the area closure the shell-count ratio is replaced by the charge-plane
area ratio, with an independently calibrated coefficient.
The linear model independently determines its own initial charge and
endpoint coefficient from the same two potentials.

\subsection{Results and numerical checks}

The nonlinear solver uses SciPy 1.15.3 adaptive collocation with relative
residual tolerance \(2\times10^{-8}\), 241 initial nonuniform mesh nodes,
and up to 30\,000 adaptive nodes. The implementation rescales the
potential by \(k_{\mathrm B}T/e\) and the liquid-side coordinate by
\(\kappa^{-1}\) for numerical conditioning; Equations~\eqref{eqS:NPB}
and \eqref{eqS:NPB_boundary} state the equivalent physical problem.
The outer boundary is 24 Debye lengths beyond the slipping plane.
Inward integration uses DOP853 with relative and absolute tolerances
\(2\times10^{-10}\) and \(2\times10^{-12}\).

At \((\kappa^{-1},\tlig,\dslip)=(3,0,1)\,\mathrm{nm}\),
\begin{align}
\frac{|Q_{\mathrm{eff}}(0)|}{4\pi\varepsilon(k_{\mathrm B}T/e)}
&=\SI{118.305635}{nm},&
\frac{|Q_{\mathrm{eff}}(\Btwo)|}{4\pi\varepsilon(k_{\mathrm B}T/e)}
&=\SI{8.793788}{nm}
\nonumber\\
\frac{|Q_{\mathrm{eff}}(\Btwo)|}{|Q_{\mathrm{eff}}(0)|}
&=0.0743311,&
\alpha_{\mathrm{NL}}&=0.299307
\label{eqS:NLnumbers}
\end{align}
The endpoint charge-plane values of
\(|e\psi(\Rs)|/(k_{\mathrm B}T)\) are 0.756792 and 0.653761.
Table~\ref{tabS:nonlinear} collects the four screening scenarios.
Constant total charge remains constant between field conditions within each
model; the linear and nonlinear models do not assume identical inferred
absolute charges. In the \(\rho_Q\) column, the zero-field nonlinear
charge is first determined from \(\zeta_0\), then retained at the
smaller endpoint particle while the potential is solved again:
\begin{equation}
\rho_Q
=\frac{|\zeta(B_2;\,Q_{\mathrm{eff}}=Q_{\mathrm{eff}}(0))|}
{|\zeta_0|}
\label{eqS:constant_charge_ratio}
\end{equation}
Thus \(\rho_Q>1\) denotes a predicted increase in zeta magnitude at
constant total charge, opposite to the measured ratio
\(\rho_{\mathrm{obs}}<1\). This is a separate comparison from the
shell-linked calculation, in which the total charge is allowed to change.

\begin{table}[htbp]
\centering
\caption{Fixed \(\tlig=0\), \(\dslip=\SI{1}{nm}\) scenarios.
The coefficients reproduce the endpoint ratio by construction. The
constant-charge column is a separate forward nonlinear prediction.}
\label{tabS:nonlinear}
\begin{adjustwidth}{-\extralength}{0cm}
\centering
\begin{tabular}{cccc}
\toprule
\(\kappa^{-1}\) (nm) & Linear shell \(\alpha\) & Nonlinear shell \(\alpha\) & Nonlinear \(\rho_Q\)\\
\midrule
1.5 & 0.143863 & 0.115753 & 3.42050\\
3.0 & 0.316017 & 0.299307 & 4.83474\\
5.0 & 0.513419 & 0.499170 & 5.50528\\
8.0 & 0.756647 & 0.743650 & 5.76280\\
\bottomrule
\end{tabular}
\end{adjustwidth}
\end{table}

Inverse-to-forward reproduction errors are below
\(1.2\times10^{-9}\,\mathrm{mV}\) across the four scenarios.
For the reference calculation, the remote boundary was changed from 24 to
16 and 32 Debye lengths, the initial mesh doubled to 481 nodes, and all
relevant tolerances tightened tenfold. The largest relative charge change was
\(1.4\times10^{-11}\), the largest absolute shell-coefficient change
\(2.1\times10^{-11}\), and the largest absolute constant-charge ratio
change \(2.8\times10^{-11}\). Boundary residuals in those convergence
tests were below \(6.7\times10^{-24}\) in the implementation's scaled units.
Weak-potential inverse and forward checks at \SI{0.01}{mV} agree with
the analytic linear charge and potential within \(2.9\times10^{-8}\)
relative difference.

The convergence tests assess the numerical solution under the stated physical
assumptions. Ion-specific and soft-corona effects remain outside this model.
The nonlinear constant-charge trend applies to the conditions evaluated here;
a general monotonicity result has not been established.

\section{Applicability and Reproducibility}

The linear solution estimates the charge-plane thermal-potential magnitude as
\begin{equation}
\frac{|e\psi(\Rs)|}{k_{\mathrm B}T}
=\frac{e|\zeta|}{k_{\mathrm B}T}
\frac{\Rs+\dslip}{\Rs}\exp(\kappa\dslip)
\label{eqS:chi_zeta}
\end{equation}
This dimensionless diagnostic compares electrostatic and thermal energies;
linearization requires it to be small compared with one.
The three inverse-map slices use the maximum of this diagnostic across the
two observed endpoint potentials. They share the full computed coefficient
range and hatch points where either endpoint diagnostic is at least one.
This hatching is a warning about the weak-potential assumption, not a
calibrated error boundary. The separate zero-field and \SI{250}{mT} maps
retain \(\dslip\) as an explicit axis and show the individual endpoint
diagnostics on one logarithmic color scale. The four nonlinear benchmarks
quantify the approximation only at their specified parameter points.
The parameter \(\kappa r\) compares diffuse-layer thickness with particle
radius; interparticle overlap additionally requires separation information.

The accompanying Python code includes the inputs, solver settings,
convergence tests, and numerical results. Population histograms are calculated
by integrating the initial size-bin probabilities through the structural and
electrokinetic relations. The frequencies, statistics, and calibration
conditions are given in Section S9.

\section{Population Propagation and Empirical Comparisons}

This section derives the population potentials from the measured size
frequencies and compares charge coefficients determined using measured and
structurally predicted endpoint sizes.

\subsection{Size frequencies and population statistics}

Table~\ref{tabS:frequencies} lists the digitized
size-frequency histograms associated with the published silver-particle
data \cite{Kthiri2021,Tawalbeh2025}. The statistical calculation assigns each bin's frequency to its center.
Frequencies are normalized by their sums, 99.9 and 100.0. These digitized
frequencies are not raw particle counts, and their decimal precision does
not specify measurement uncertainty. 

\begin{table}[htbp]
\centering
\caption{Digitized size frequencies. Blank entries indicate the absence
of a corresponding histogram bin.}
\label{tabS:frequencies}
\begin{adjustwidth}{-\extralength}{0cm}
\centering
\begin{tabular}{rrrr}
\toprule
\multicolumn{2}{c}{0 mT} & \multicolumn{2}{c}{250 mT}\\
\(r\) (nm) & Frequency & \(r\) (nm) & Frequency\\
\midrule
12 & 0.9 & 2 & 2.55\\
13 & 1.95 & 3 & 6.6\\
14 & 0.9 & 4 & 12.5\\
15 & 3.6 & 5 & 19\\
16 & 4.6 & 6 & 17.25\\
17 & 7.4 & 7 & 14.55\\
18 & 8.45 & 8 & 11\\
19 & 12.25 & 9 & 6.95\\
20 & 15.95 & 10 & 5.5\\
21 & 10.4 & 11 & 3\\
22 & 8.45 & 12 & 1.1\\
23 & 6.4 &  & \\
24 & 5.6 &  & \\
25 & 4.6 &  & \\
26 & 2.6 &  & \\
27 & 1.95 &  & \\
28 & 1.95 &  & \\
29 & 1.95 &  & \\
\midrule
Total & 99.9 & Total & 100.0\\
\bottomrule
\end{tabular}
\end{adjustwidth}
\end{table}

Population index \(j=1\) denotes \(B_1=0\), and \(j=2\) denotes
\(B_2=\SI{250}{mT}\). The size-bin centers are \(r_{ji}\).
Writing the digitized frequency of bin \(i\) as \(f_{ji}\), its
normalized probability is
\begin{equation}
w_{ji}=\frac{f_{ji}}{\sum_k f_{jk}}
\qquad
\sum_iw_{ji}=1
\label{eqS:frequency_weights}
\end{equation}
The bin frequency \(f_{ji}\) is distinct from the field-response function
\(f(B)\). For example, the zero-field \SI{20}{nm} bin has
\(w_{1i}=15.95/99.9\), whereas the field-prepared \SI{5}{nm} bin has
\(w_{2i}=19/100\). These probabilities come from the measured size
frequencies, not from the structural trajectory or an assumed functional
form for the size distribution. Number weighting means that a bin
contributes according to its fraction of nanoparticles, without an
additional factor of particle area, volume, or scattering intensity.

For values \(v_i\) with normalized weights \(w_i\), the population
moments are
\begin{equation}
\langle v\rangle=\sum_iw_i v_i,\qquad
\sigma_v=\left[\sum_iw_i(v_i-\langle v\rangle)^2\right]^{1/2},
\qquad \mathrm{RSD}_v=\frac{\sigma_v}{|\langle v\rangle|}
\label{eqS:population_stats}
\end{equation}
These expressions describe the discrete population rather than uncertainty
in an estimated mean. The symbol \(\sigma_v\) here is a
statistical standard deviation, distinct from the charge density
\(\sigma_{\mathrm{eff}}\). With no raw count or replicate information,
a standard error or confidence interval is not assigned.

The weighted cumulative distribution and its quantile are
\begin{equation}
F_v(z)=\sum_iw_i\mathbf{1}_{v_i\leq z},\qquad
q_p=\inf\{z:F_v(z)\geq p\}
\label{eqS:population_cdf}
\end{equation}
Here \(\mathbf{1}\) is the indicator of the stated inequality. Reported
quantiles are discrete inverse-CDF values, without interpolation.
For the potential curves, \(v_i=\zeta_i(B)\) is sorted by its signed
value before accumulating the weights. The lower and upper ribbon
boundaries use \(p=0.05\) and \(0.95\), respectively; the lower boundary
is therefore the more negative potential. The mean is the weighted sum,
not the midpoint of these bounds. Because the distribution is discrete,
a cumulative jump can carry the enclosed probability beyond 90\%.
The displayed 5th--95th percentile ribbon describes population spread,
not uncertainty in a predicted mean.

\subsection{Population potentials and charge calibration}

At fixed \(\kappa,\tlig,\dslip\), each size has
\(\Rs(r_{ji})=r_{ji}+\tlig\) and \(x_{ji}=r_{ji}/a\).
Applying the same charge law and electrostatic solution to every size gives
\begin{equation}
\zeta_{ji}
=\frac{q_0[1+\alpha_{\mathrm{pop}}f(B_j)]
\nshell(r_{ji}/a)\exp(-\kappa\dslip)}
{4\pi\varepsilon[\Rs(r_{ji})+\dslip][1+\kappa\Rs(r_{ji})]}
\label{eqS:population_kernel}
\end{equation}
The subscript \(\mathrm{pop}\) identifies the coefficient calibrated to population
means rather than to the single characteristic particle; it is not a
size-specific charging parameter. A single \(q_0\) and a single
\(\alpha_{\mathrm{pop}}\) apply across both populations. They are obtained
from the two assumed mean-potential conditions
\(\sum_iw_{1i}\zeta(r_{1i},0)=\zeta_0\) and
\(\sum_iw_{2i}\zeta(r_{2i},B_2)=\zeta_B\).
The first fixes \(q_0\): at zero field the factor
\(1+\alpha_{\mathrm{pop}}f(0)\) equals one, so the measured initial
mean determines the common charge amplitude. The initial predicted
potential assigned to each radius is then obtained from
Equation~\eqref{eqS:population_kernel}; no individual-bin potential
is measured or independently fitted. The second condition fixes the
single field-dependent multiplier using the measured final size
frequencies. Dividing the second mean condition by the first eliminates
the common amplitude and leaves
\(1+\alpha_{\mathrm{pop}}\), because \(f(B_2)=1\):
\begin{align}
q_0&=
\frac{4\pi\varepsilon\zeta_0}
{\displaystyle\exp(-\kappa\dslip)
\sum_iw_{1i}
\frac{\nshell(r_{1i}/a)}
{[\Rs(r_{1i})+\dslip][1+\kappa\Rs(r_{1i})]}}
\nonumber\\
1+\alpha_{\mathrm{pop}}
&=\rho_{\mathrm{obs}}
\frac{\displaystyle\sum_iw_{1i}
\frac{\nshell(r_{1i}/a)}
{[\Rs(r_{1i})+\dslip][1+\kappa\Rs(r_{1i})]}}
{\displaystyle\sum_iw_{2i}
\frac{\nshell(r_{2i}/a)}
{[\Rs(r_{2i})+\dslip][1+\kappa\Rs(r_{2i})]}}
\label{eqS:population_fit}
\end{align}
Equating the reported zeta potentials with number-weighted means is an
assumption of the population calculation. Number-weighted theoretical potentials cannot in general be
identified with an instrument-weighted mobility estimate or its converted
zeta value \cite{Delgado2005}. No instrument response or mobility
distribution is reconstructed by Equation~\eqref{eqS:population_fit}.

The discrete predicted potential distribution assigns the original weight
\(w_{ji}\) to \(\zeta_{ji}\). For the reference interface the coefficient is
0.196860. The size and predicted-potential moments are listed in
Table~\ref{tabS:population_moments}. Because the charge weight is common
and the potential sign does not change, the relative width satisfies
\begin{equation}
\mathrm{RSD}_{\zeta,j}
=\left[
\frac{\displaystyle\sum_iw_{ji}
\left\{\frac{\nshell(r_{ji}/a)}
{[\Rs(r_{ji})+\dslip][1+\kappa\Rs(r_{ji})]}\right\}^{\!2}}
{\displaystyle\left\{\sum_iw_{ji}
\frac{\nshell(r_{ji}/a)}
{[\Rs(r_{ji})+\dslip][1+\kappa\Rs(r_{ji})]}\right\}^{\!2}}
-1\right]^{1/2}
\label{eqS:population_cv}
\end{equation}
To obtain Equation~\eqref{eqS:population_cv}, one substitutes
Equation~\eqref{eqS:population_kernel} into
\(\sigma_\zeta^2=\langle\zeta^2\rangle-\langle\zeta\rangle^2\)
and divides by \(|\langle\zeta\rangle|^2\). At a fixed field,
\(q_0[1+\alpha_{\mathrm{pop}}f(B)]\exp(-\kappa\dslip)/(4\pi\varepsilon)\)
is common to every size, so it cancels from this relative width.
Consequently, changing the common charge multiplier rescales the mean
and the absolute standard deviation by the same magnitude, but leaves
their ratio unchanged. Heterogeneous charge weights or size-dependent interfacial parameters would
require a joint distribution of size and interfacial properties.
The size dependence is proportional to
\(\nshell(r/a)/[(r+\tlig+\dslip)(r+\tlig+\kappa^{-1})]\).
The exact two-length-sum degeneracy therefore survives this linear population
mapping when \(q_0\) is free; adding size polydispersity does not itself
identify all three lengths.

\begin{table}[htbp]
\centering
\caption{Discrete population moments at the reference interface.
The experimental zeta values are taken as the mean potentials of the
measured size populations when determining the charge parameters. The transported population retains that calibration without refitting.
SD denotes population standard deviation, not standard error.}
\label{tabS:population_moments}
\begin{adjustwidth}{-\extralength}{0cm}
\centering
\begin{tabular}{lrrrr}
\toprule
Population & Mean \(r\) (nm) & SD \(r\) (nm) & Mean \(\zeta\) (mV) & SD \(\zeta\) (mV)\\
\midrule
Empirical 0 mT & 20.403 & 3.500 & -13.200 & 0.488\\
Empirical 250 mT & 6.270 & 2.204 & -10.000 & 2.173\\
Transported to 250 mT & 5.09573 & 0.83515 & -9.13055 & 0.99423\\
\bottomrule
\end{tabular}
\end{adjustwidth}
\end{table}

\subsection{Preserved-weight structural propagation}

The structural population relation applies the endpoint ratio of \(x/n'(x)\)
to each initial size \cite{Tawalbeh2025,Tawalbeh2026APB}. With \(x_i(B)=r_i(B)/a\) and
\(x_{1i}=r_{1i}/a\), this gives
\begin{equation}
\frac{x_i(B)}{n'[x_i(B)]}
=\left[1+(\lambda-1)\left(\frac{B}{\Btwo}\right)^2\right]
\frac{x_{1i}}{n'(x_{1i})},
\qquad
\lambda=\frac{x_2/n'(x_2)}{x_1/n'(x_1)}=4.206878
\label{eqS:population_flow}
\end{equation}
The dimensionless ratio \(\lambda\) gives the same fractional change in
\(x/n'(x)\) for every initial size. Thus each trajectory is determined by
the common structural relation rather than a separate fit.
At \(B=0\) the multiplier is one and the physical branch returns
\(r_{1i}\). At \(B=B_2\) it equals \(\lambda\).
For the particular initial radius \(r_{1i}=r_1=\SI{20}{nm}\), this
condition returns \(r_i(B_2)=r_2=\SI{5}{nm}\).
It does not require the mean of an initial distribution to follow that
single characteristic trajectory, since the radius transformation is
nonlinear. The measured zero-field mean is \SI{20.403}{nm}, not the
characteristic input \SI{20}{nm}.
The explicit derivative in Equation~\eqref{eqS:nderivative} again gives
a quadratic, whose larger root is selected. The implementation checks a
nonnegative discriminant, \(r_i(B)>a\delta\), and preservation of support
ordering throughout the reported calculation.

Each point retains its initial probability \(w_{1i}\). The transported mean
potential is therefore
\begin{equation}
\overline{\zeta}_{\mathrm{tr}}(B)
=\frac{q_0[1+\alpha_{\mathrm{pop}}(B/\Btwo)^2]
\exp(-\kappa\dslip)}{4\pi\varepsilon}
\sum_iw_{1i}
\frac{\nshell[r_i(B)/a]}
{[\Rs(r_i(B))+\dslip][1+\kappa\Rs(r_i(B))]}
\label{eqS:population_transport}
\end{equation}
Here \(\Rs(r_i(B))=r_i(B)+\tlig\), with the same \(q_0\) and
\(\alpha_{\mathrm{pop}}\) determined from the two measured size
populations. Preserving each initial probability is an assumption relating
separately prepared populations; it does not imply that individual particles
are tracked or that their experimental numbers are conserved. The endpoint
histograms alone do not establish this correspondence.

The transported endpoint support is
\([3.08675,7.14499]\,\mathrm{nm}\), whereas the empirical endpoint spans
2--\SI{12}{nm}. To quantify this discrepancy in radius units, the
one-dimensional Wasserstein distance is calculated directly from the weighted
discrete cumulative distributions:
\begin{equation}
W_1
=\int_{-\infty}^{\infty}
\left|F_{\mathrm{tr}}(r)-F_{2}^{\mathrm{emp}}(r)\right|\,dr
=\SI{1.41370}{nm}
\label{eqS:population_wasserstein}
\end{equation}
Here \(F_{\mathrm{tr}}\) and \(F_2^{\mathrm{emp}}\) are the endpoint
transported and empirical size CDFs defined by the rule in
Equation~\eqref{eqS:population_cdf}. Between adjacent radii in the combined
set of both populations, each CDF is constant. The integral is therefore
evaluated by multiplying the absolute difference of the two CDF levels
by that interval's width and summing over intervals. \(W_1\) measures
the average radius displacement required to match two normalized size
populations under an optimal redistribution of probability. Its unit is
nanometres, and it is zero only when the two discrete distributions are
identical. This is a descriptive mismatch, not a significance test. The transported mean radius and width are lower than the
empirical values (Table~\ref{tabS:population_moments}). With the unchanged
empirical coefficient, its mean potential is \SI{-9.13055}{mV}.
Recalibrating against the transported rather than empirical endpoint would
instead give \(\alpha_{\mathrm{pop}}=0.310829\). The distinction can be
made explicit by dividing the transported endpoint mean by the measured
endpoint mean while keeping \(q_0\) and the measured-size coefficient fixed:
\begin{align}
\frac{\overline{\zeta}_{\mathrm{tr}}(B_2)}{\zeta_B}
&=
\frac{\displaystyle\sum_iw_{1i}
\frac{\nshell[r_i(B_2)/a]}
{[\Rs(r_i(B_2))+\dslip][1+\kappa\Rs(r_i(B_2))]}}
{\displaystyle\sum_iw_{2i}
\frac{\nshell(r_{2i}/a)}
{[\Rs(r_{2i})+\dslip][1+\kappa\Rs(r_{2i})]}}
\nonumber\\
&=\frac{1.1968597739}{1.3108294044}=0.913055
\label{eqS:population_endpoint_mismatch}
\end{align}
The denominator averages the potential's size dependence over the
measured field-prepared population; the numerator averages it over the
structurally predicted population with the initial weights. These are
different distributions, so their averages need not agree. Equation~
\eqref{eqS:population_endpoint_mismatch} gives
\(\overline{\zeta}_{\mathrm{tr}}(B_2)
=(-10.0)\times0.913055=\SI{-9.13055}{mV}\).
This is the separation between the mean curve and the \SI{250}{mT}
experimental symbol in Figure 6d, also shown in
Figure~\ref{figS:population_transport}c.

Both population coefficients differ from the characteristic-particle
value \(\alpha=0.316017\), which was determined from the single radius
pair in Equation~\eqref{eqS:Tworked}. The distinction is the population
used in the calibration, not a change in the field law or the physical
meaning of the coefficient. Replacing 0.196860 with 0.310829 would
impose the measured endpoint mean on the structurally predicted
population; it would not improve the predicted size distribution.
This second calibration is compared with the fixed-coefficient result
in Section S9.5. Nucleation,
selection, aggregation, or other size-dependent processes could change the
population weights, but the observed difference does not identify which
process is responsible.

\subsection{Small-size tail, applicability, and reproducibility}

The \SI{2}{nm} empirical endpoint bin has probability 2.55\%
and is included in the measured-size calculations. Its contribution
to the conditional endpoint mean is \SI{-0.09888}{mV}.
Continuum electrostatics and a common sharp interface are particularly
uncertain for this small-size tail. Its influence is assessed by restricting
the endpoint to \(r\geq\SI{3}{nm}\) and renormalizing the remaining
weights after excluding 2.55\% of the population. The selected threshold
is a sensitivity condition, not an established validity boundary. With the
original calibration fixed, the conditional mean is
\SI{-10.16020}{mV} and SD \SI{1.95909}{mV}. If the
conditional subset were separately recalibrated to \SI{-10}{mV}, its
coefficient would be 0.177988. This alternative is not substituted
for the full-population result.

Across all retained empirical bins at the reference interface, the maximum
of \(|e\psi(\Rs)|/(k_{\mathrm B}T)\) is 0.814, with no mass at or above
unity. This is not small compared with one and does not establish quantitative
accuracy. The screening scan records the maximum diagnostic and the weight
of bins with a diagnostic at least one at every screening value; its
small-screening portion includes such bins. The nonlinear
characteristic-particle benchmarks in Section S7 are not an error validation
for every population bin. Instrument response, ion-specific effects, and the
extrapolative small-size tail remain separate issues.

\begin{figure}[!ht]
\begin{adjustwidth}{-\extralength}{0cm}
\centering
\includegraphics[width=\linewidth]{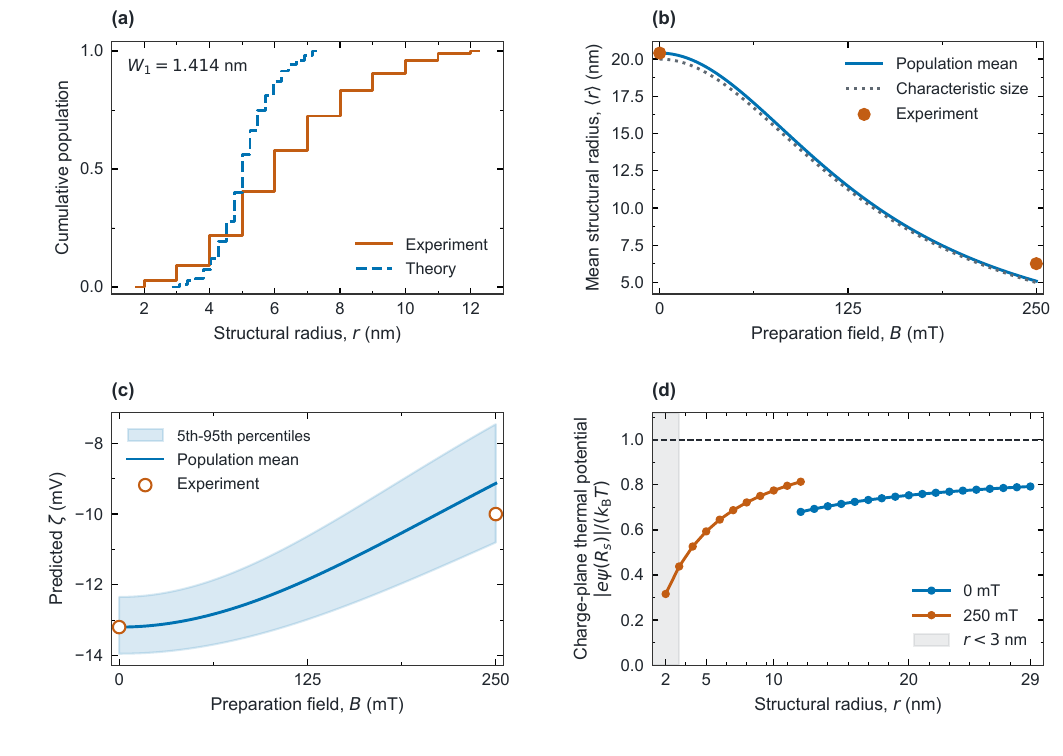}
\end{adjustwidth}
\caption{Population predictions and linear-electrostatic applicability.
(\textbf{a}) Measured and predicted endpoint size CDFs; the annotated
\(W_1=\SI{1.414}{nm}\) is their one-dimensional Wasserstein distance
from Equation~\eqref{eqS:population_wasserstein}.
(\textbf{b}) Predicted mean radius, characteristic trajectory, and measured
endpoint means. The orange markers are means of the measured size
distributions, not the characteristic size points in Figure 1a.
(\textbf{c}) Predicted mean potential and 5th--95th
population percentiles, with the charge coefficient determined from the
measured size populations. Open symbols are experimental zeta values,
assumed to represent number-weighted means; the band shows population
spread, not uncertainty in the mean. (\textbf{d}) Charge-plane thermal
potential for the measured size-bin centers (markers); the connecting
lines guide the eye. All calculations use
\((\kappa^{-1},\tlig,\dslip)=(3,0,1)\,\mathrm{nm}\).
The shaded \(r<\SI{3}{nm}\) region identifies the small-size sensitivity
test, not a universal cutoff; the unit line marks failure of the
weak-potential condition rather than a specified numerical error.}
\label{figS:population_transport}
\end{figure}

In Figure~\ref{figS:population_transport}a, each CDF step is the
accumulated nanoparticle fraction up to the radius on the horizontal
axis, and the printed \(W_1\) is the discrepancy defined above.
Panel b distinguishes the number-weighted predicted mean from the
\(\SI{20}{nm}\)-to-\(\SI{5}{nm}\) characteristic trajectory; the
endpoint symbols are the means of the measured size histograms.
Panel c uses the same charge calibration, mean, and quantile definitions
as Figure 6d. In panel d, each marker is the calculated charge-plane
thermal-potential magnitude for one measured size-bin center.
The connecting lines guide the eye; they are not additional
measurements. The dashed unit level marks where the potential energy
equals \(k_{\mathrm B}T\), and the shaded small-radius region identifies
the sensitivity test described above.

Numerical checks verify probability normalization, recovery of the initial
sizes at zero field, the structural endpoint conditions, the two mean-potential
constraints, and the invariance of relative width under a common charge
multiplier. The two-length-sum invariance also holds for the calibrated
population distribution. Independent calculations confirm the population
moments, shell relation, and discrete CDF distance. The zeta histograms use
shared \SI{0.5}{mV} intervals; reported statistics are calculated from the
original size-bin centers. Inputs and code are supplied with the manuscript.

\subsection{Structural endpoint solution and potential distributions}

To construct the endpoint histogram from the initial size population, we
solve Equation~\eqref{eqS:population_flow} at \(B=B_2\) for each
initial \(x_{1i}=r_{1i}/a\). The positive branch is \cite{Tawalbeh2025}
\begin{align}
x_i(B_2)={}&\frac{n'(x_{1i})}{6\phi_{\mathrm b}\lambda x_{1i}}
\Bigg\{1-\frac{6\delta(\phi_{\mathrm d}-\phi_{\mathrm b})\lambda x_{1i}}
{n'(x_{1i})}\nonumber\\
&\quad+\left[
\left(1-\frac{6\delta(\phi_{\mathrm d}-\phi_{\mathrm b})\lambda x_{1i}}
{n'(x_{1i})}\right)^2
+\frac{36\delta^2\phi_{\mathrm b}(\phi_{\mathrm d}-\phi_{\mathrm b})
\lambda^2x_{1i}^2}{[n'(x_{1i})]^2}
\right]^{1/2}\Bigg\}
\label{eqS:eq15_explicit}
\end{align}
Here \(\lambda\) is the same structural endpoint ratio in
Equation~\eqref{eqS:population_flow}. The resulting endpoint radius is
\(r_i(B_2)=a x_i(B_2)\), retaining each initial probability
\(w_{1i}\). The implementation checks the real discriminant, the
large-radius branch, and the relation
\(x_i(B_2)/n'[x_i(B_2)]=\lambda x_{1i}/n'(x_{1i})\).

\begin{figure}[!htbp]
\begin{adjustwidth}{-\extralength}{0cm}
\centering
\includegraphics[width=\linewidth]{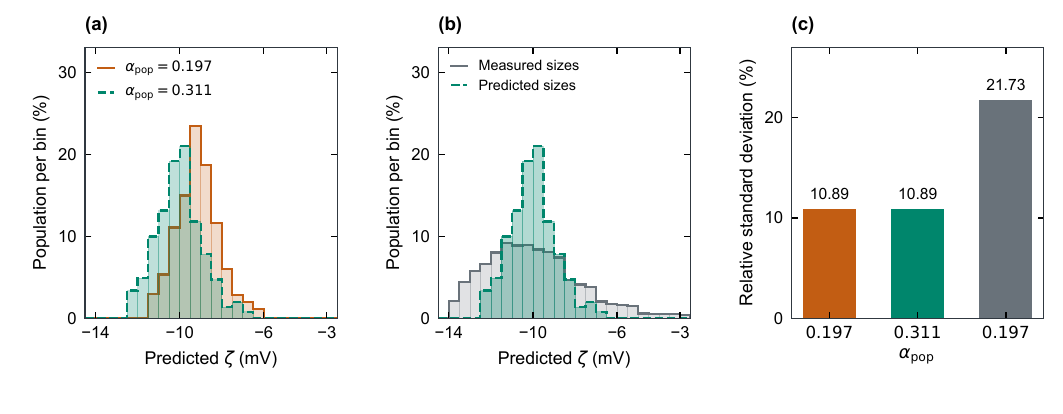}
\end{adjustwidth}
\caption{Predicted endpoint zeta distributions at
\((\kappa^{-1},\tlig,\dslip)=(3,0,1)\,\mathrm{nm}\).
(\textbf{a}) Structurally predicted populations with the measured-size charge
coefficient \(\alpha_{\mathrm{pop}}=0.196860\) (orange, solid) and
with \(\alpha_{\mathrm{pop}}=0.310829\) determined by imposing a mean
of \SI{-10}{mV} (green, dashed).
(\textbf{b}) Predicted potentials from theoretical (green) and measured
(gray) endpoint sizes, each calibrated to the experimental mean.
The gray histogram is calculated from the measured sizes; it is not a
measured zeta distribution.
(\textbf{c}) Relative population standard deviations,
\(100\sigma_\zeta/|\langle\zeta\rangle|\): predicted sizes with
\(\alpha_{\mathrm{pop}}=0.196860\) (orange) and 0.310829 (green),
and measured sizes with \(\alpha_{\mathrm{pop}}=0.196860\) (gray).
Structural propagation retains the initial size weights; changing the charge
coefficient leaves the size distribution unchanged. The \SI{0.5}{mV} bins
show population percentages and each histogram sums to 100\%.
Calibration and statistics use the original bin centers. All zeta
distributions are calculated, and their widths describe population spread
rather than measurement uncertainty.}
\label{figS:eq15_calibration}
\end{figure}

We then apply Equation~\eqref{eqS:population_kernel} to these predicted
sizes. The calculation with fixed charge parameters uses the zero-field \(q_0\)
and \(\alpha_{\mathrm{pop}}=0.1968597739\) from
Equation~\eqref{eqS:population_fit}, giving a mean of
\SI{-9.13055}{mV} and SD \SI{0.99423}{mV}. Although its final sizes
are calculated from the initial distribution and structural endpoint
conditions, its charge parameters are determined using both measured size
populations. For comparison, keeping \(q_0\) fixed but calibrating
the transported endpoint mean to \SI{-10}{mV} gives
\(\alpha_{\mathrm{pop}}=0.3108294044\) and SD
\SI{1.08890}{mV}. Algebraically, this follows from
Equation~\eqref{eqS:population_fit} by replacing only its final-size
average, \(\sum_iw_{2i}(\cdots)\), with the average over
\(r_i(B_2)\) and \(w_{1i}\). The initial-size average and \(q_0\)
are unchanged. Relative to the original calculation, every endpoint
potential and its absolute SD are multiplied by
\(1.3108294044/1.1968597739=1.095224\). This alternative imposes its endpoint mean and does
not validate the structural prediction. Neither calculation adjusts the
transported size weights. Figure~\ref{figS:eq15_calibration} compares
these alternatives with the separately mapped empirical endpoint, whose
center-based potential SD is \SI{2.17278}{mV}.

For histogram construction, probability is taken to be uniform within each
initial interval \(r_{1i}\pm\SI{0.5}{nm}\). We propagate
this mass through the composition of Equation~\eqref{eqS:eq15_explicit}
and the potential map. To write the probability calculation explicitly, let
\(r_{1i}^{\pm}=r_{1i}\pm\SI{0.5}{nm}\) denote the original size-bin
edges and let \(\zeta_k^-\) and \(\zeta_k^+\) denote the edges of a
potential bin. Write \(r(B;r_0)\) for the structural solution starting
at an arbitrary radius \(r_0\). The probability assigned to potential
bin \(k\) is
\begin{equation}
P_k(B)=\sum_i
\frac{w_{1i}}{r_{1i}^{+}-r_{1i}^{-}}
\int_{r_{1i}^{-}}^{r_{1i}^{+}}
\mathbf{1}_{\{
\zeta_k^-\leq\zeta[r(B;r_0),B]<\zeta_k^+
\}}\,dr_0
\label{eqS:histogram_probability}
\end{equation}
The function \(\zeta[r(B;r_0),B]\) is the composition of the
structural relation and Equation~\eqref{eqS:population_kernel}; it is
not a new charge law. Inverting this monotonic composition at the edges
of shared \SI{0.5}{mV} zeta bins determines which part of each initial
interval satisfies the indicator. Its length divided by
\(r_{1i}^{+}-r_{1i}^{-}\), multiplied by \(w_{1i}\), gives the
contribution of that initial bin. This is equivalent to evaluating
Equation~\eqref{eqS:histogram_probability} by interval overlap.

At zero field \(r(0;r_0)=r_0\), so the same construction gives the
initial potential histogram. For the empirical endpoint comparison,
one instead uses the measured endpoint intervals, weights \(w_{2i}\),
and their direct potential map, without structural propagation.
For the size histogram, the indicator selects a final radius interval
rather than a potential interval. Propagation retains the nonuniform
density produced within each final radius interval; treating that final
interval as uniform would in general give different potential-bin
probabilities.

A plotted potential-bin height is \(100P_k\), the percentage of
nanoparticles whose predicted zeta potential lies in that interval.
It is not a probability density per millivolt. Every initial probability
is allocated once, giving \(\sum_kP_k=1\) when the displayed bins cover
the full mapped support. Thus each histogram sums to 100\%.
Calibrations and reported statistics use the original bin centers,
whereas histogram bars use the stated within-bin reconstruction.
The mean estimated from the centers of the plotted potential bins need
not reproduce the center-based calibration exactly. The original size-bin centers give
an endpoint support of
\([3.08675,7.14499]\,\mathrm{nm}\), with no mass below
\SI{3}{nm}; the inferred initial edges instead map to
\([2.96667,7.26406]\,\mathrm{nm}\), placing approximately
0.1250\% of the reconstructed mass below \SI{3}{nm}. This difference
results from the within-bin representation; the original weights are unchanged.

Figure~\ref{figS:eq15_calibration}a compares the two charge calibrations
on the same structurally predicted endpoint sizes. The displacement and
change in absolute width result solely from their different common
charge multipliers. Panel b compares the predicted-size and
measured-size mappings after each has been constrained to the same
mean. Their unequal widths remain because their size populations differ.
Panel c reports \(100\,\mathrm{RSD}_\zeta\), not histogram probability:
the two structurally predicted populations both have relative SD
10.89\%, whereas the measured-size mapping has relative SD 21.73\%.
The equal first two bars follow directly from
Equation~\eqref{eqS:population_cv}. All three potential distributions are
calculated from size information; no measured distribution of
individual-particle zeta potentials is available for a direct width
comparison.

\reftitle{References}
\bibliography{ref}
 \end{document}